\documentclass[pre,aps,twocolumn,showpacs]{revtex4}
\usepackage[dvips]{graphicx}

\begin{document}

\title{Conformation dependent electronic transport in a 
DNA double-helix}

\author{Sourav Kundu}
\email{sourav.kundu@saha.ac.in}
\affiliation{Condensed Matter Physics Division, 
Saha Institute of Nuclear Physics, 
1/AF, Bidhannagar, Kolkata 700 064, India}

\author{S. N. Karmakar}
\email{sachindranath.karmakar@saha.ac.in}
\affiliation{Condensed Matter Physics Division, 
Saha Institute of Nuclear Physics, 
1/AF, Bidhannagar, Kolkata 700 064, India}

\begin{abstract}
 In this work we report the study of conformation dependent 
electronic transport properties of DNA double-helix within 
tight-binding framework including its helical symmetry. We 
have studied the changes in localization properties of DNA 
as we alter the number of stacked bases within a pitch of the 
double-helix keeping the total number of nucleotides in the DNA 
chain fixed. We take three DNA sequences, two of them are periodic 
and one is random and observe that localization length increases 
as we increase the radius of DNA double-helix {\it i.e.}, number 
of nucleotides within a pitch. We have also investigated the 
effect of backbone energetic on the I-V response of the system and 
we find that in presence of helical symmetry, depending on the 
interplay of conformal variation and disorder strength DNA can be 
found in either metallic or semiconducting and even in an insulating 
phase, which in turn successfully explain all the experimental findings 
by a single model.

\end{abstract}


%\keywords{DNA, localization, conformation, electronic transport} 

\maketitle

\section{Introduction}
 The advancements of nanoscience and technology with everyday 
encouraging a growing number of scientists across the various 
disciplines to devise ingenious ways for decreasing the size 
and increasing the performance of the nano-electronic circuits. 
One of the promising route is to use molecules and molecular 
structures as a component of those circuits. From these efforts 
a new branch has emerged called molecular electronics. Among 
different branches of molecular electronics, DNA and alike 
biomolecules have drawn maximum attention in the last decade 
from both the theoreticians as well as experimentalists and 
still growing in numbers. The main reason behind this attraction 
is the potential of DNA to become an inevitable agent for the 
future nanoelectronic devices and computers, as it might can serve 
in different ways in a nano-electronic circuits such as a wire, transistor 
or a switch depending on its electronic properties~\cite{endres, dekker}. 
Not only this, a precise knowledge of charge transfer mechanism through 
DNA could help in understanding the process like oxidative damage sensing, 
protein binding, gene regulation and cell division. On the other hand 
electrical properties, specially conductivity of DNA can be used for 
marker-free gene test~\cite{kleine} which is one of the most highly desired 
biophysical methods~\cite{mckendry}. Inspite of the vast efforts from physicists 
as well as biologists around the world, charge transport results through DNA are 
still quite controversial~\cite{fink, porath, cai, tran, zhang, storm, yoo}. 
Experimentally it is found that DNA can behave either as a good conductor~\cite{fink}, 
semiconductor~\cite{porath, yoo}, insulator~\cite{storm, pablo} and even as 
a superconductor~\cite{kasumov} at low temperature. Several experiments both 
on synthetic periodic DNA chains~\cite{porath, yoo} as well as unordered 
sequence of basepairs~\cite{cohen, hihtah} show the presence of a conduction 
gap in I-V curves at room temperature. Whereas linear response observed in 
Ref.~\cite{xu} and both the staircase and linear behaviour in I-V curves shown 
in poly(dG)-poly(dC) chains~\cite{hwang}. 
%Thus reproducible experimental 
%results are still a great technical challenge due to complex structure of DNA 
%molecule, environmental effects, thermal vibration and contact resistance variation. 
Due to this experimental ambiguity and lack of understanding of 
charge transfer mechanism in DNA, leads to different phenomenological models 
in which charge transfer is mediated via polarons~\cite{conwell}, 
solitons~\cite{hermon} or electrons or holes~\cite{dekker,ratner,beratan}. 

 This diversity of experimental findings on transport properties of DNA 
is due to several reasons such as, DNA varies widely in terms of its 
composition, length and structure, presence of counterions and impurities 
which can attach to the phosphate group of the backbones, environmental 
effects, thermal vibration and contact resistance variation.
In this communication, we try to address the effects of structure of DNA 
{\it i.e.}, conformal behaviour on its transport properties. Experiments 
done more than half a century ago by Wilkins {\it et. al.},~\cite{wilkins} 
first suggested that overstretched DNA (quite longer than its natural length) 
undergoes transition to a structure that can accommodate elongation up 
to twice the length of a relaxed DNA. Crucial developments in understanding 
mechanical properties of DNA was achieved via stretching experiments~\cite{smith, 
cluzel, strick}. Depending on the stretching force applied, DNA first uncoils, 
then exhibit stiff elastic response and at last undergoes an abrupt structural 
transformation. Now as all the DNA are twisted (natural double-helix structure)  
and the amount of twist-stretch {\it i.e.}, radius of the helix varies 
from one situation to another, this study has to be made in details. People 
have already tried to study the effects of conformation introducing twist angle 
or chirality~\cite{yega, song} into {\it ab-initio} calculations. Studies 
also have been done on electronic properties of stretched DNA~\cite{maragakis} 
but the effects of helical structure and conformality on its transport properties 
are yet not well explored. While study within much simpler tight-binding framework 
is hardly available in current literature. In our work we try to find out these 
effects within the tight-binding model. To do this we follow Ref.~\cite{gore}, 
where a mechanical model of DNA is proposed. DNA being modelled as an elastic 
rod, wrapped helically by a stiff wire. The radius of elastic rod can change 
upon stretching with a Poisson's ratio $\eta$=0.5. The outer wire is affixed 
to the rod helically with a given pitch. As stretching force being applied, 
the elastic rod elongates in the length and its radius decreases. As a result 
the stiff wire overwinds and the number of turn increases. We take this mechanical 
model and interpret in the language of tight-binding formulation. We use twisted 
ladder model~\cite{sourav}, to imitate this mechanical model which includes both 
the helical symmetry and conformation. We have been able to show three 
different phases of DNA {\it i.e.}, metallic, semiconducting and insulating 
depending on helical symmetry, conformation (twist-stretching) and disorder. 
We have also found some structural configurations at which system hardly 
disturbed by external changes.

This paper is organized in the following way: In Sec. II we discuss 
about our theoretical formulation and describe the model Hamiltonian. 
We explain our numerical results in Sec. III and summarized in Sec. IV.

\section{Model and Theoretical Formulation}

DNA, carrier of genetic code of all forms of life, a $\pi$-stacked array 
of four different nitrogenous bases adenine (A), guanine (G), cytosine (C) 
and thymine (T) attached among themselves via hydrogen bond following 
complementary base pairing and coupled with sugar-phosphate backbones 
forming the double-helix structure. In most of the theoretical models, 
electronic conduction~\cite{guti2,cuni,zhong,bakhshi,ladik} is assumed 
through the long-axis of the DNA molecule. To model DNA, in our present study, 
we take the tight-binding (TB) dangling backbone ladder model~\cite{klotsa,gcuni} 
and add extra hopping channels due to the proximity of bases in the upper 
strand with the corresponding bases of the lower strand in the next pitch 
to incorporate its helical symmetry. %(see Fig.~\ref{fig1}).  
The Hamiltonian for the said model can be expressed as (for schematic 
representation of this model we refer to~\cite{sourav})

\begin{equation}
 H_{DNA}=H_{ladder}+ H_{helicity}+H_{backbone}~,
\label{hamilton}
\end{equation}

\noindent
where, 
\begin{eqnarray}
& H_{ladder}&= \sum\limits_{i=1}^N\sum\limits_{j=I,II}\left(\epsilon_{ij}
c^\dagger_{ij}c_{ij}
+t_{ij}c^\dagger_{ij}c_{i+1j}+\mbox{H.c.} \right)\nonumber \\
&&~~~~~~~~~~~~~~~~~+ \sum_{i=1}^N v \left(c^\dagger_{iI}c_{i II}+
\mbox{H.c.} \right)~, \\
& H_{helicity}&=\sum\limits_{i=1}^N v^{\prime} 
\left(c^\dagger_{i II}c_{i+n I}+ \mbox{H.c.} \right)~, \\
& H_{backbone}&=\sum\limits_{i=1}^N\sum\limits_{j=I,II}
\left(\epsilon_i^{q(j)}c^\dagger_{i q(j)}c_{i q(j)}\right.\nonumber \\
&&~~~~~~~~~~~~~~~~~+\left.t_i^{q(j)}c^\dagger_{ij}c_{i q(j)}+
\mbox{H.c.} \right)~,
\end{eqnarray} 
where $c_{ij}^\dagger$ and $c_{ij}$ are the electron creation and annihilation 
operators at the {\it i}th nucleotide at the jth stand, $t_{ij}=$ nearest 
neighbour hopping amplitude between nucleotides along the jth branch of the 
ladder, $\epsilon_{ij}=$ on-site energy of the nucleotides, $\epsilon_{i}^{q(j)}=$ 
on-site energy of the backbone site adjacent to ith nucleotide of the 
jth strand with $q(j)=\uparrow,\downarrow$ representing the upper and 
lower strands respectively, $t_{i}^{q(j)}=$ hopping amplitude between 
a nucleotide and the corresponding backbone site, $v=$ interstrand 
hopping integral between nucleotides in two strands of ladder within 
a given pitch, $v'=$ interstrand hopping integral between neighboring 
atomic sites in the adjacent pitches which actually accounts for 
the helical structure of DNA. Here $n$ denotes the number of sites 
in each strand within a given pitch. For simplicity, we set 
$\epsilon_i^{q(j)}=\epsilon_b$, $t_{ij}=t_i$ and $t_i^{q(j)}=t_b$.
        
 To explore the transport properties of DNA, we use 
semi-infinite 1D chains as source (S) and drain (D) electrodes 
connected to alternative strands of the DNA in cross-wise fashion 
to the left and right ends respectively and the Hamiltonian of the 
entire system is given by 
$ H=H_{DNA}+H_S+ H_D + H_{tun}$. 
The explicit form of $H_S$, $H_D$  and 
$H_{tun}$ are 
\begin{eqnarray}
& H_S & =\sum\limits_{i=-\infty}^0\left(\epsilon c^\dagger_ic_i+
t c^\dagger_{i+1}c_i+\mbox{H.c.} \right)~, \\
& H_D & =\sum\limits_{i=N+1}^\infty\left(\epsilon c^\dagger_ic_i+
t c^\dagger_{i+1}c_i+\mbox{H.c.} \right)~, \\
& H_{tun}& = \tau \left(c^\dagger_0c_1+c^\dagger_Nc_{N+1} +
\mbox{H.c.}\right)~,
\end{eqnarray}
where $\tau$ is the tunneling matrix element between DNA and the electrodes.         

\begin{figure*}
\centering

 \begin{tabular}{ccc}
 
  \includegraphics[width=48mm, height=35mm]{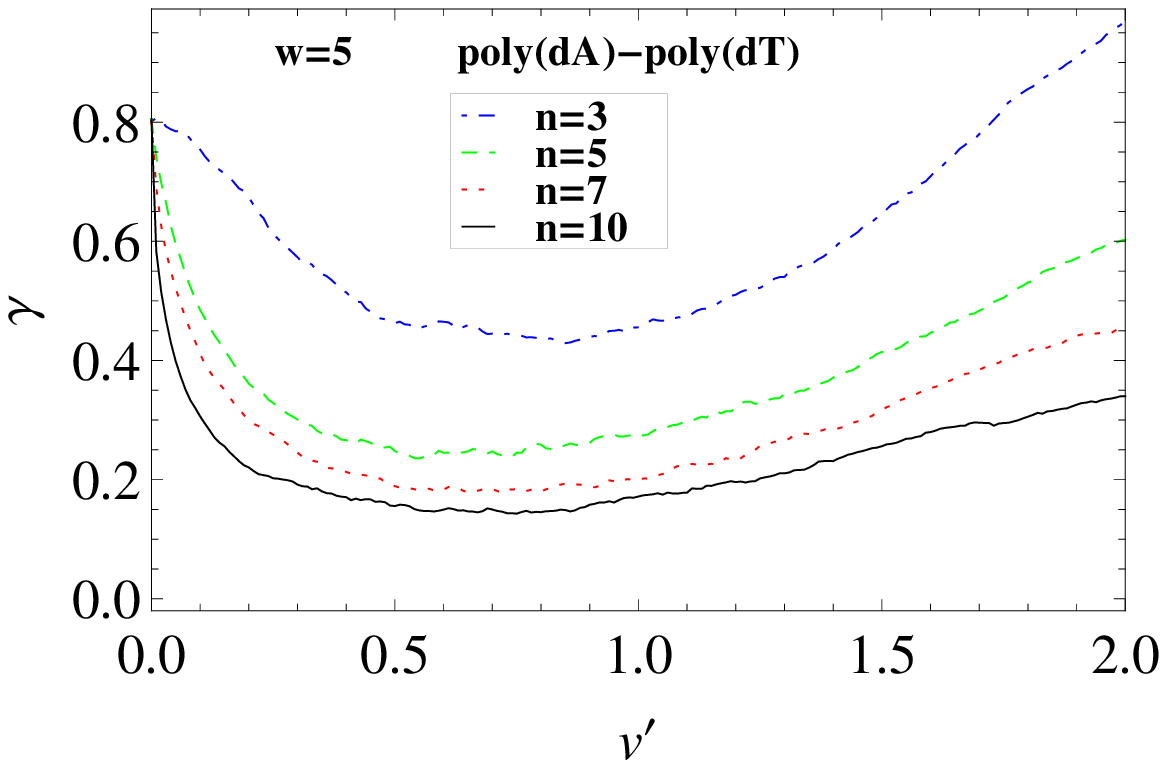}&
   \includegraphics[width=48mm, height=35mm]{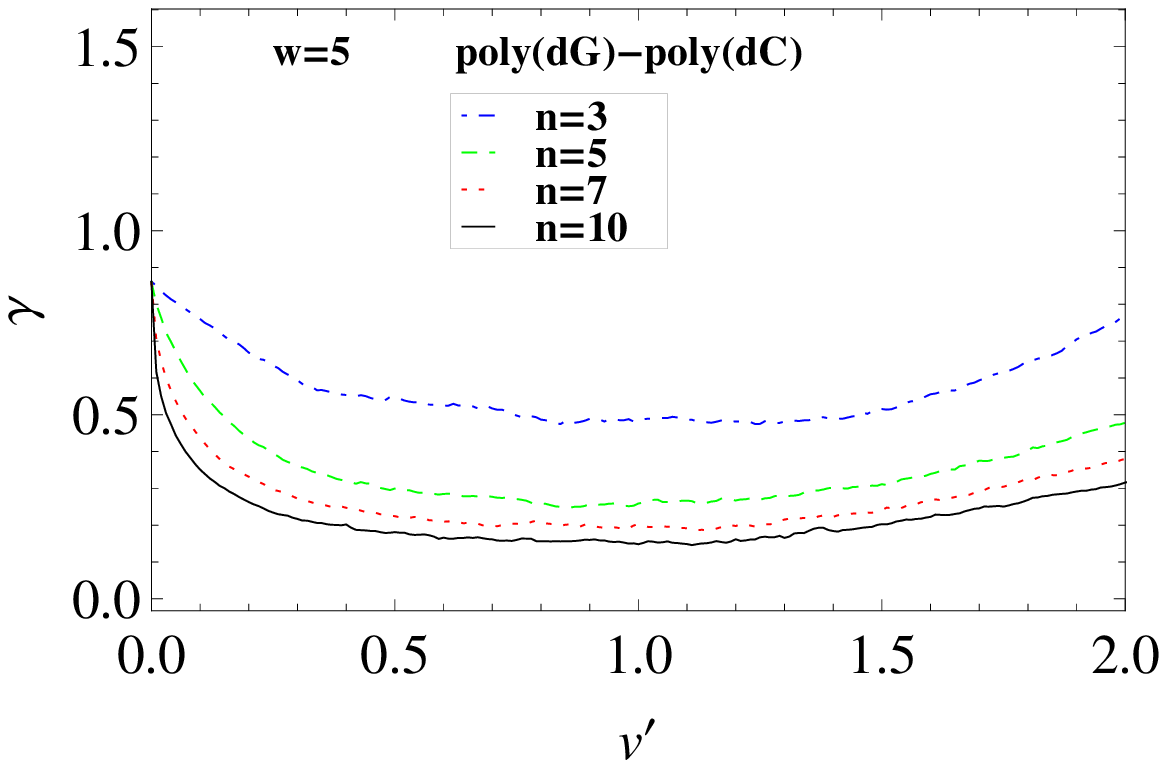}&
    \includegraphics[width=48mm, height=35mm]{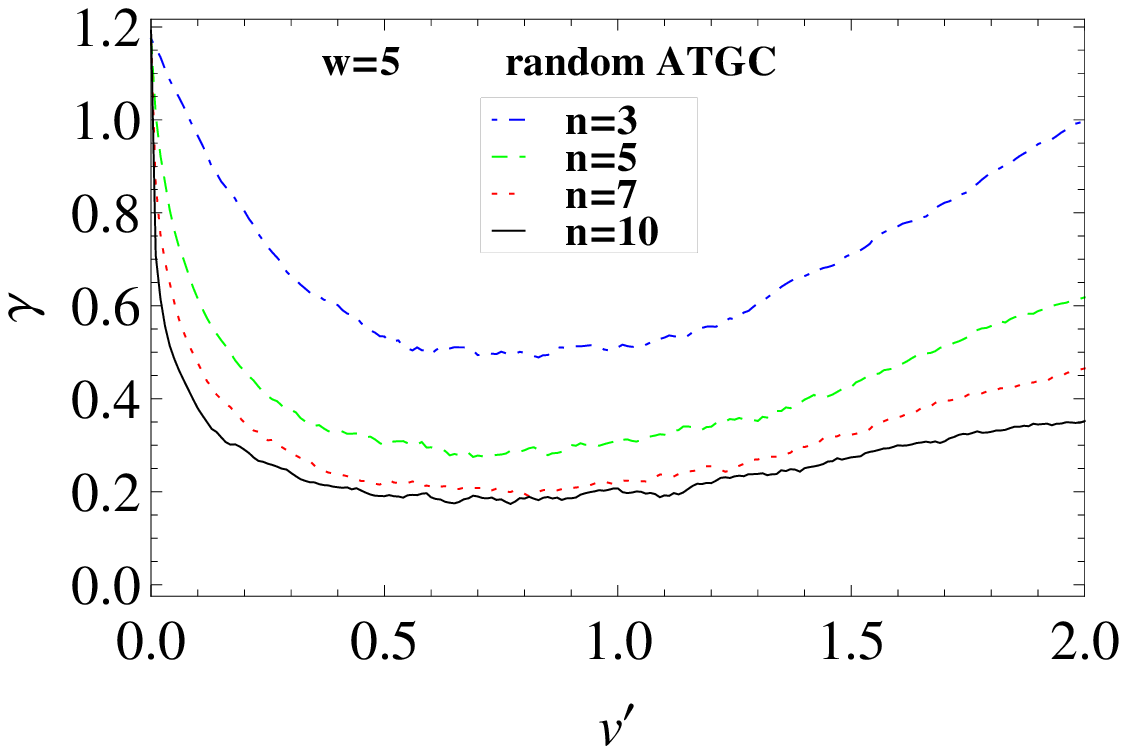}\\
    
   \includegraphics[width=48mm, height=35mm]{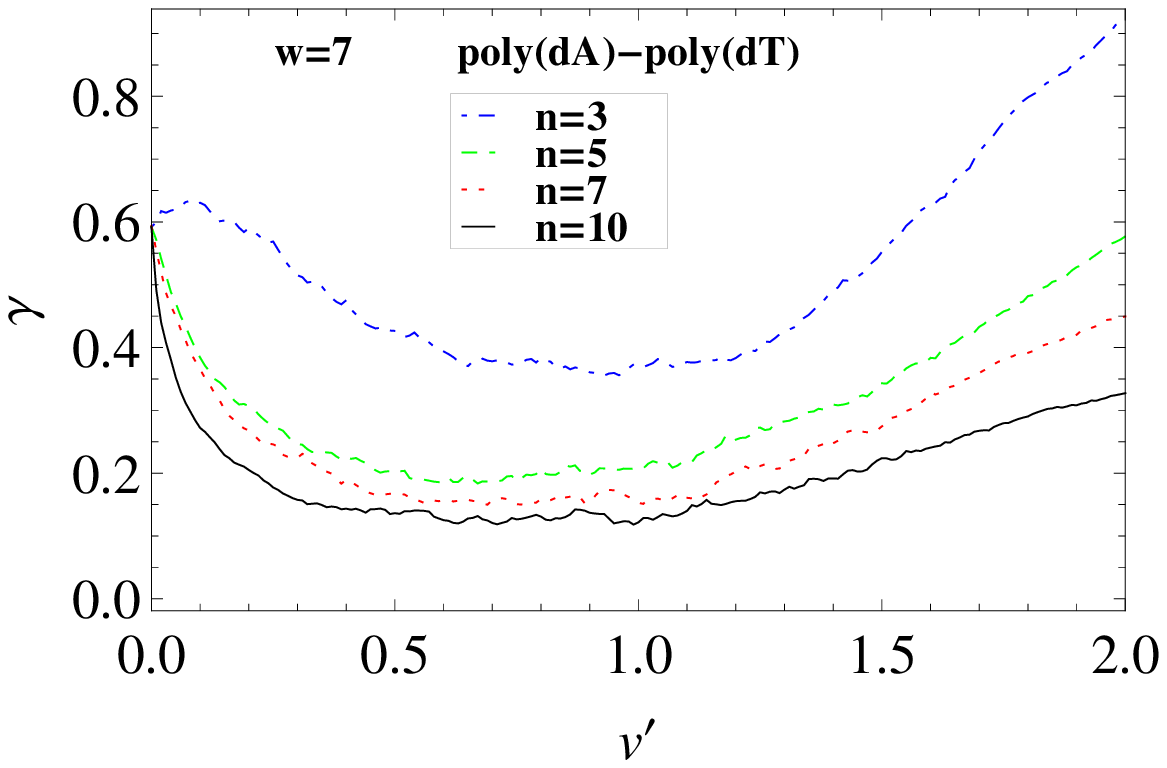}&
    \includegraphics[width=48mm, height=35mm]{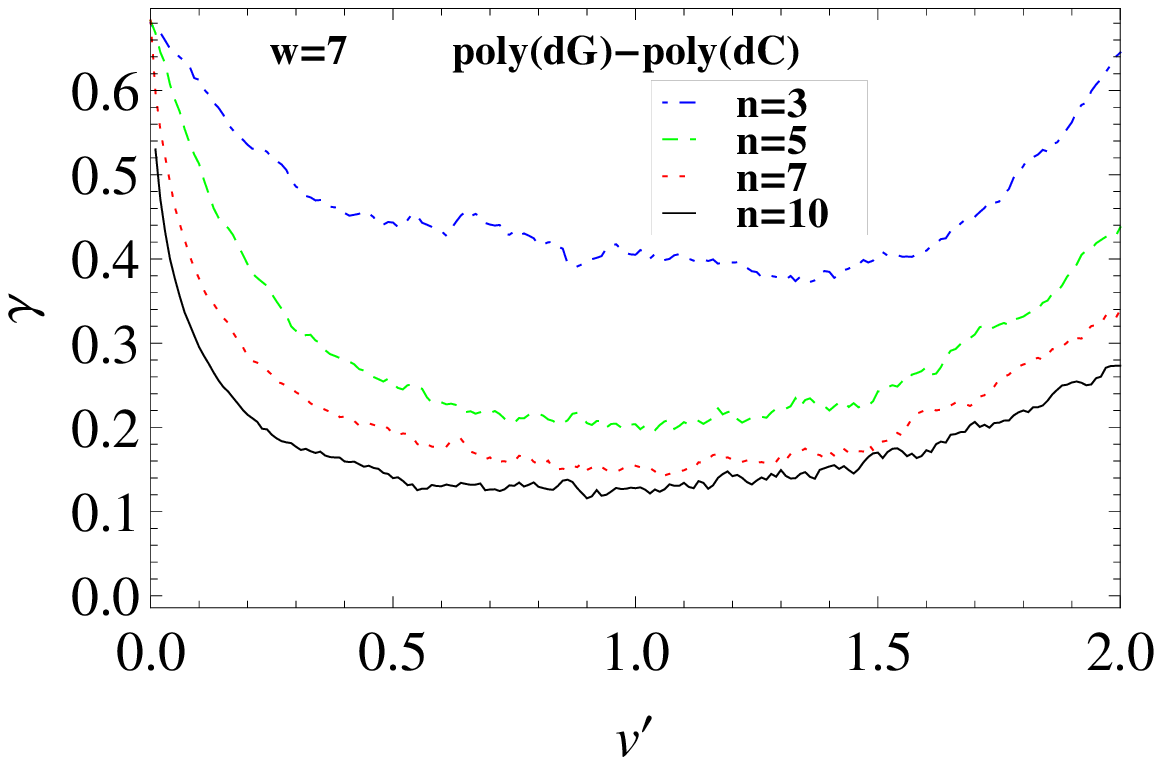}&
     \includegraphics[width=48mm, height=35mm]{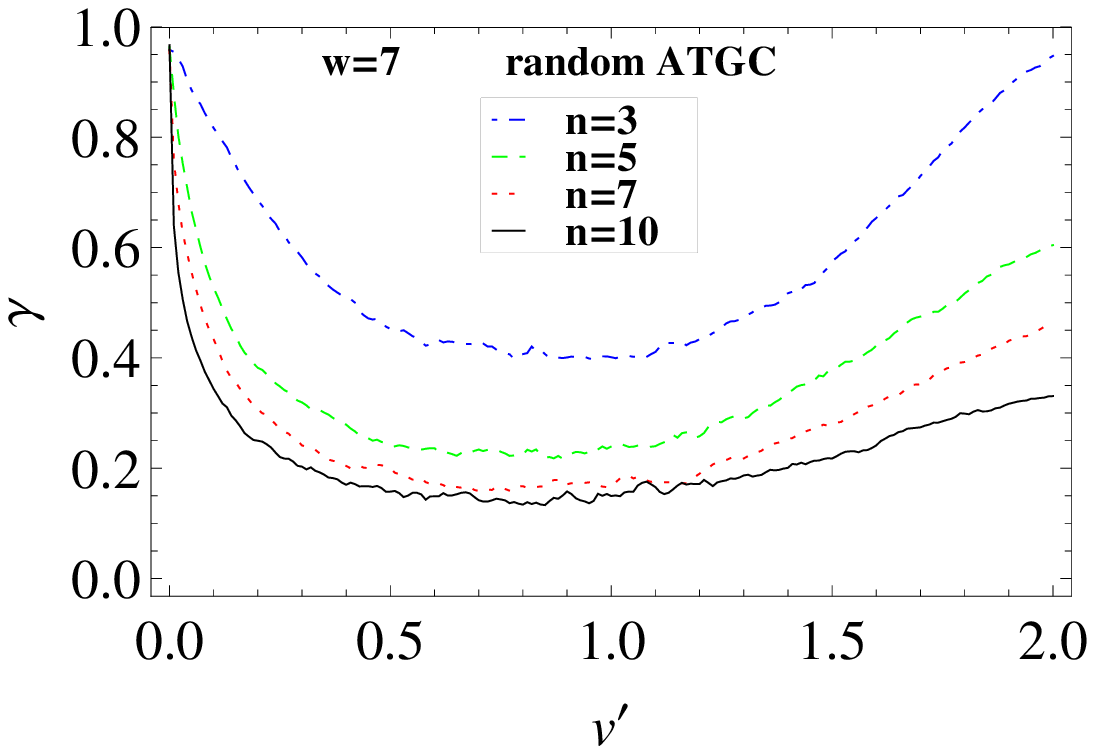}\\

    \includegraphics[width=48mm, height=35mm]{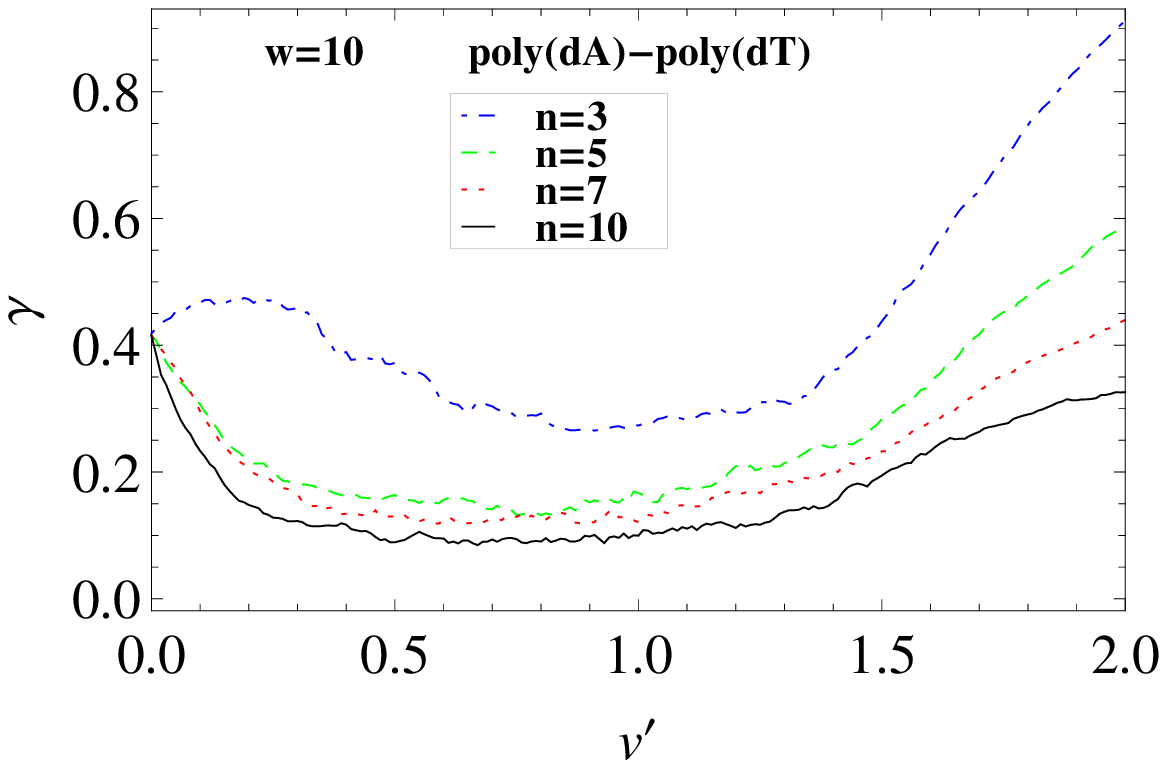}&
     \includegraphics[width=48mm, height=35mm]{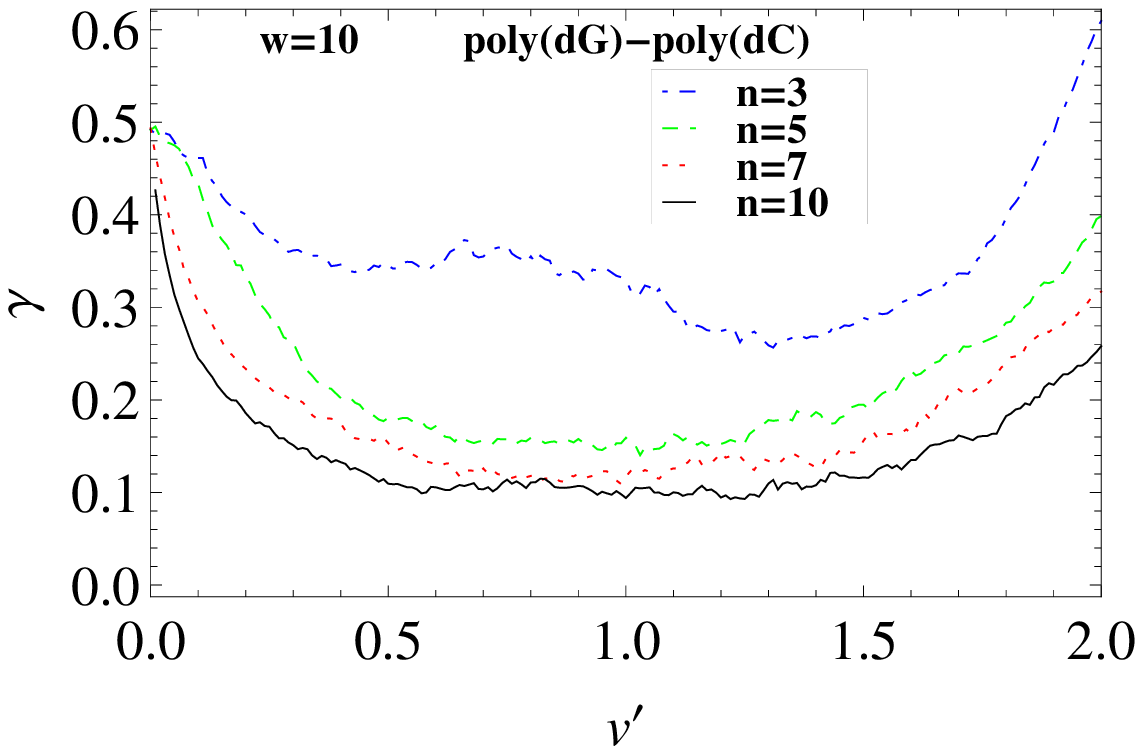}&
      \includegraphics[width=48mm, height=35mm]{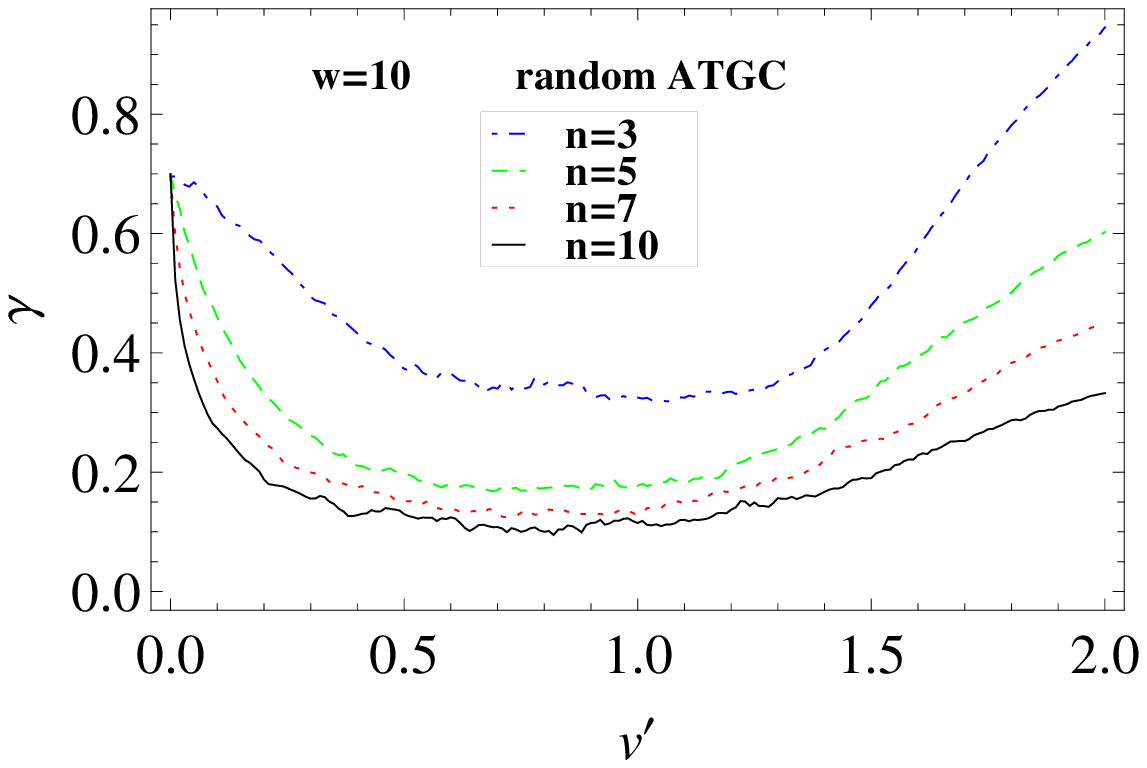}\\

     \includegraphics[width=48mm, height=35mm]{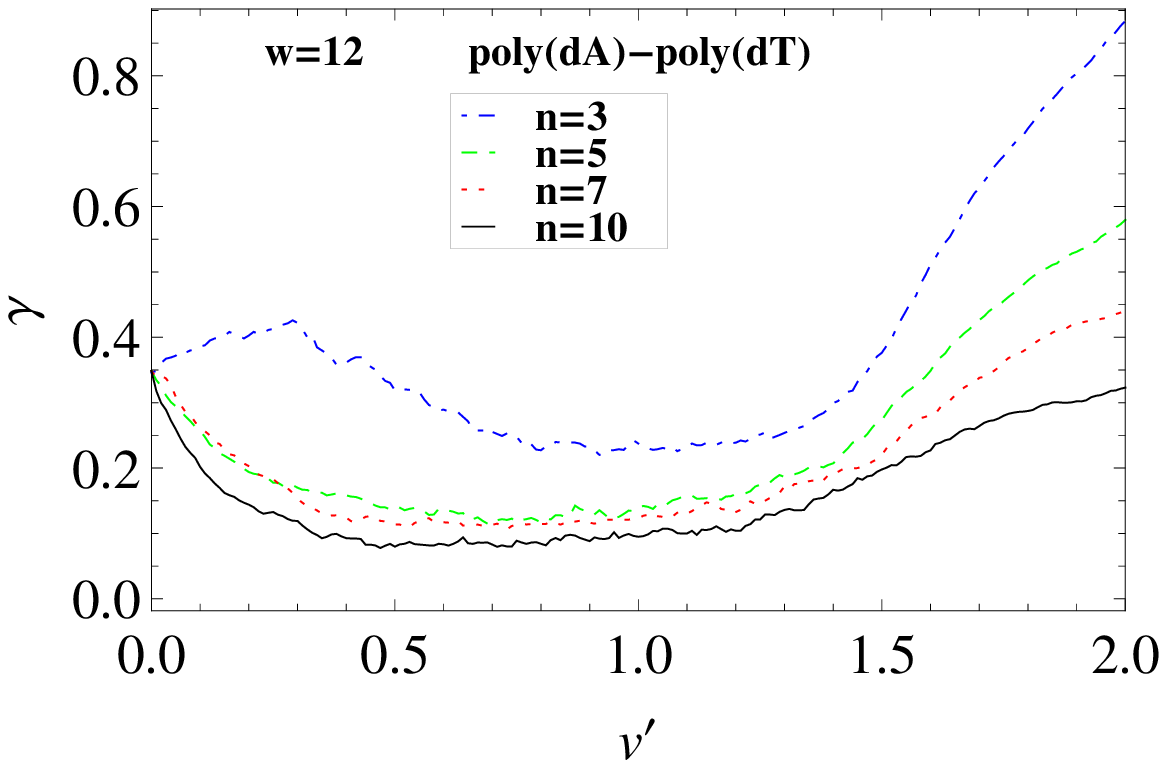}&
      \includegraphics[width=48mm, height=35mm]{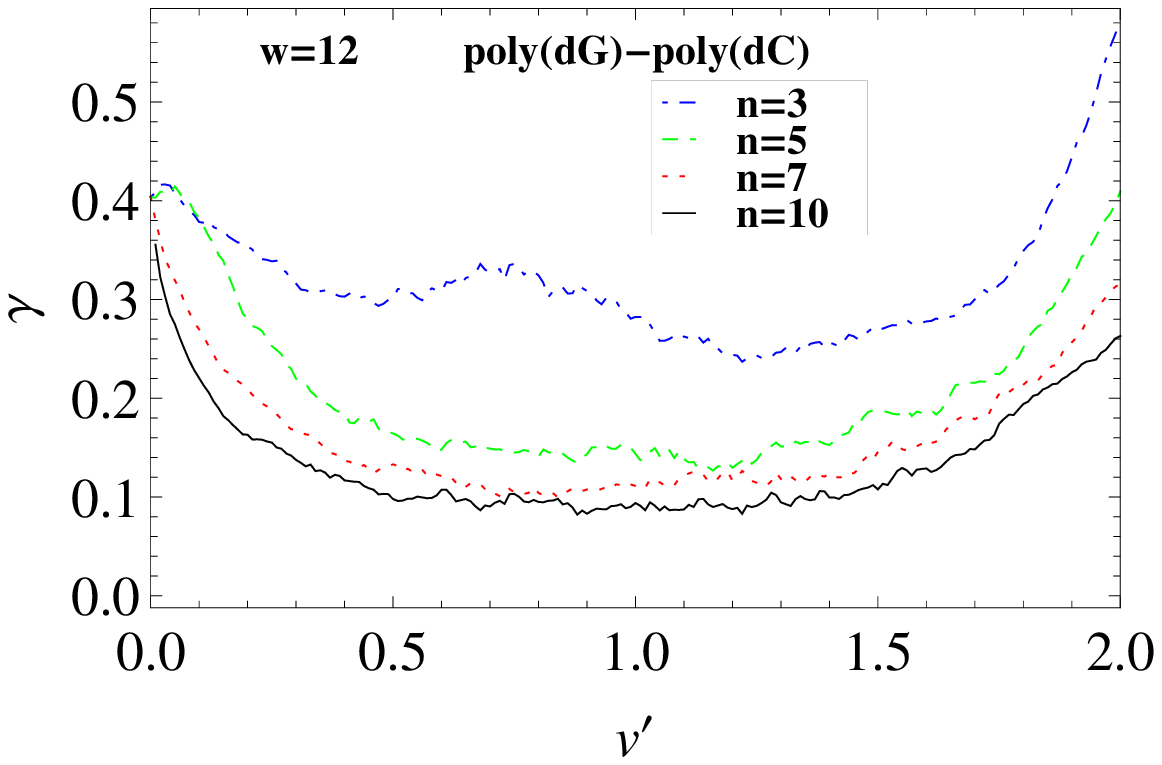}&
       \includegraphics[width=48mm, height=35mm]{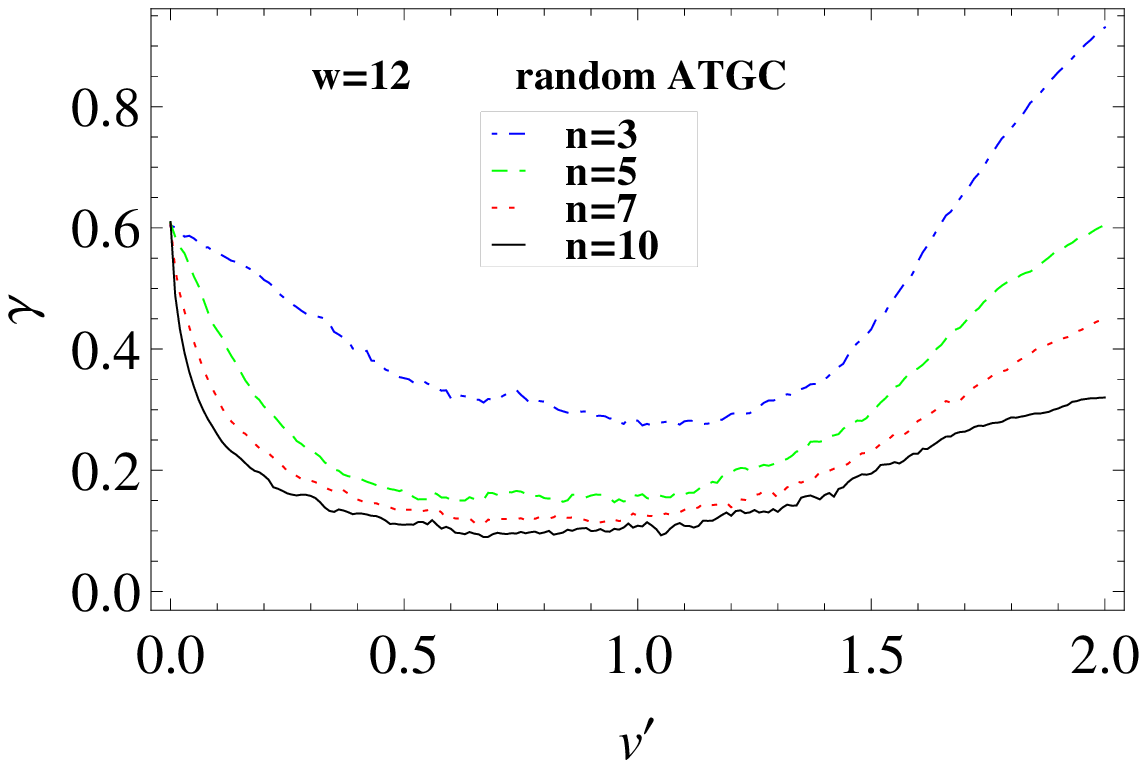}\\

      \includegraphics[width=48mm, height=35mm]{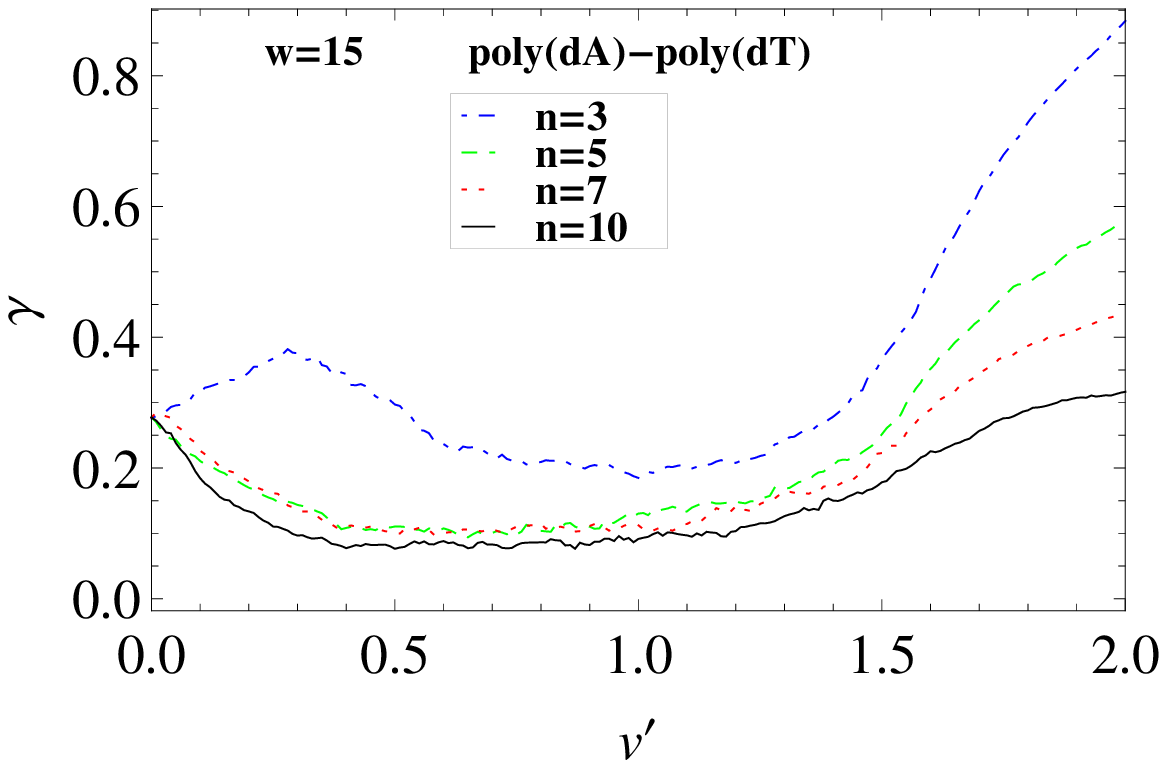}&
       \includegraphics[width=48mm, height=35mm]{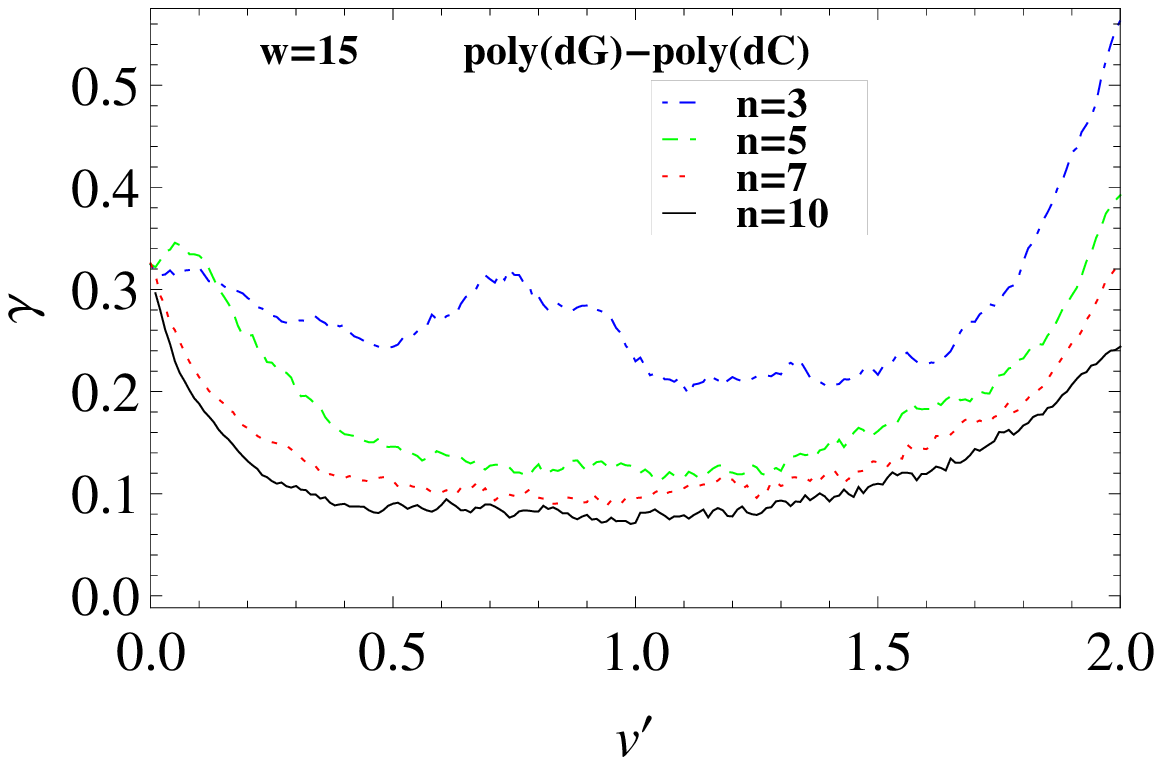}&
        \includegraphics[width=48mm, height=35mm]{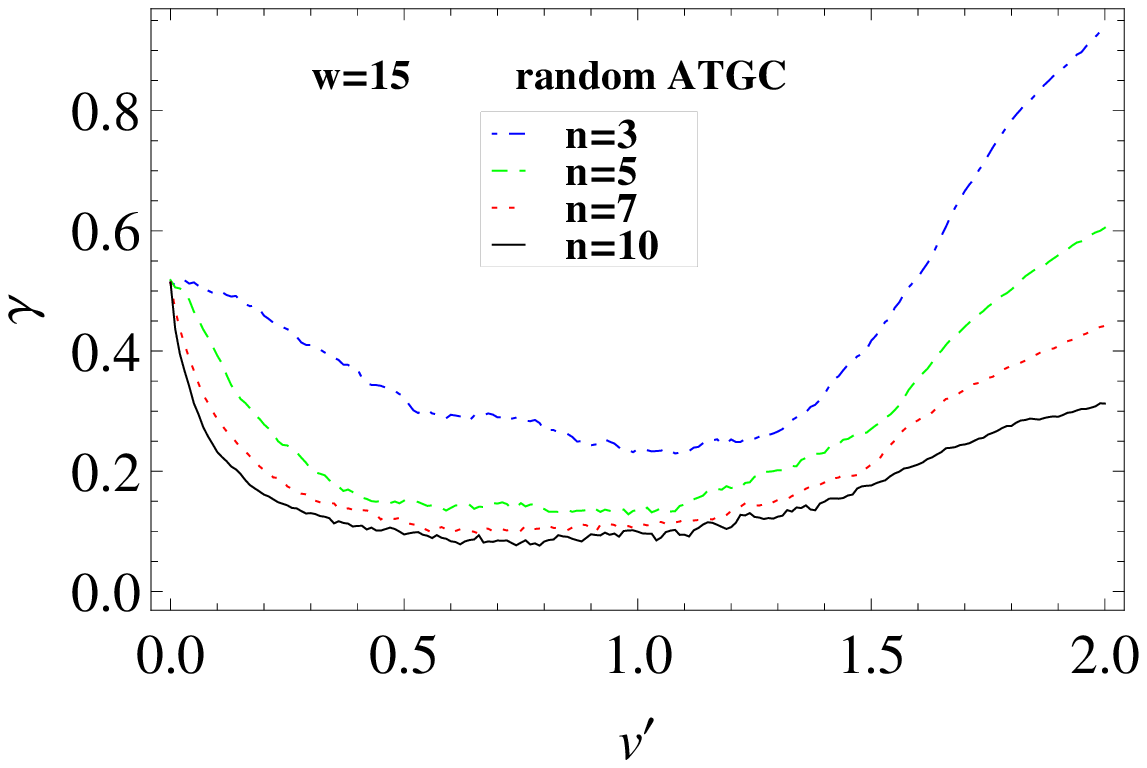}\\

       \includegraphics[width=48mm, height=35mm]{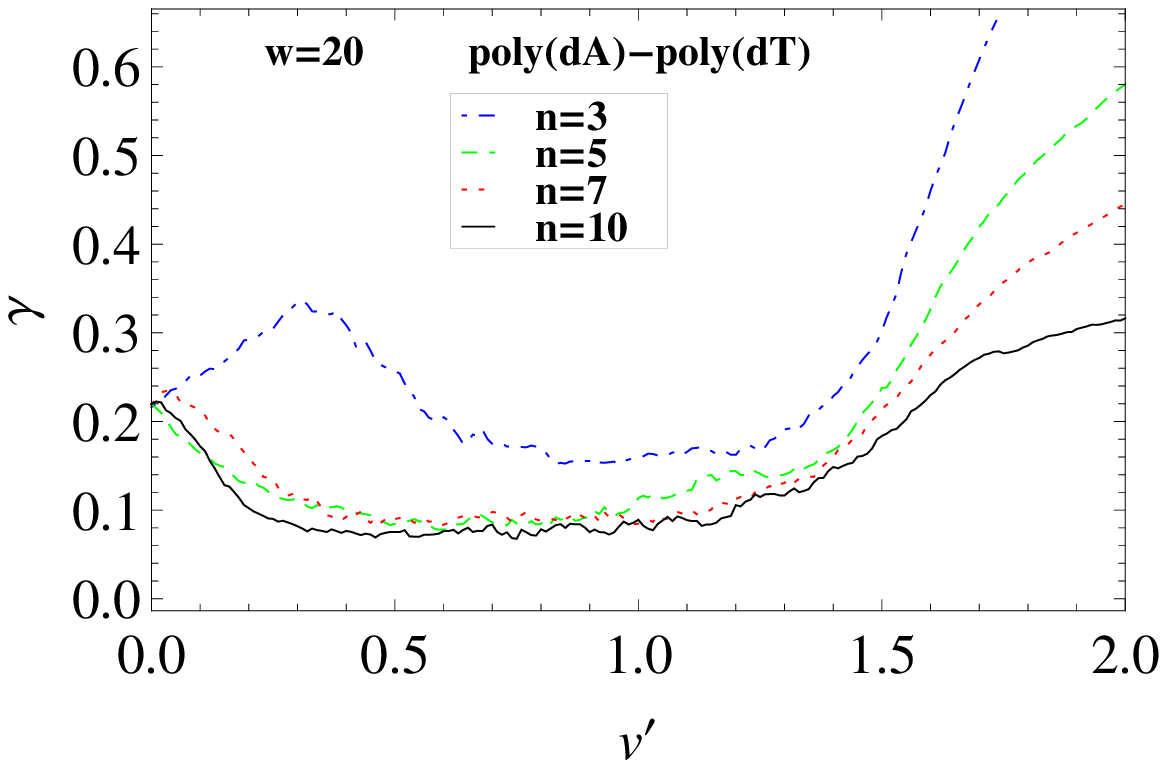}&
        \includegraphics[width=48mm, height=35mm]{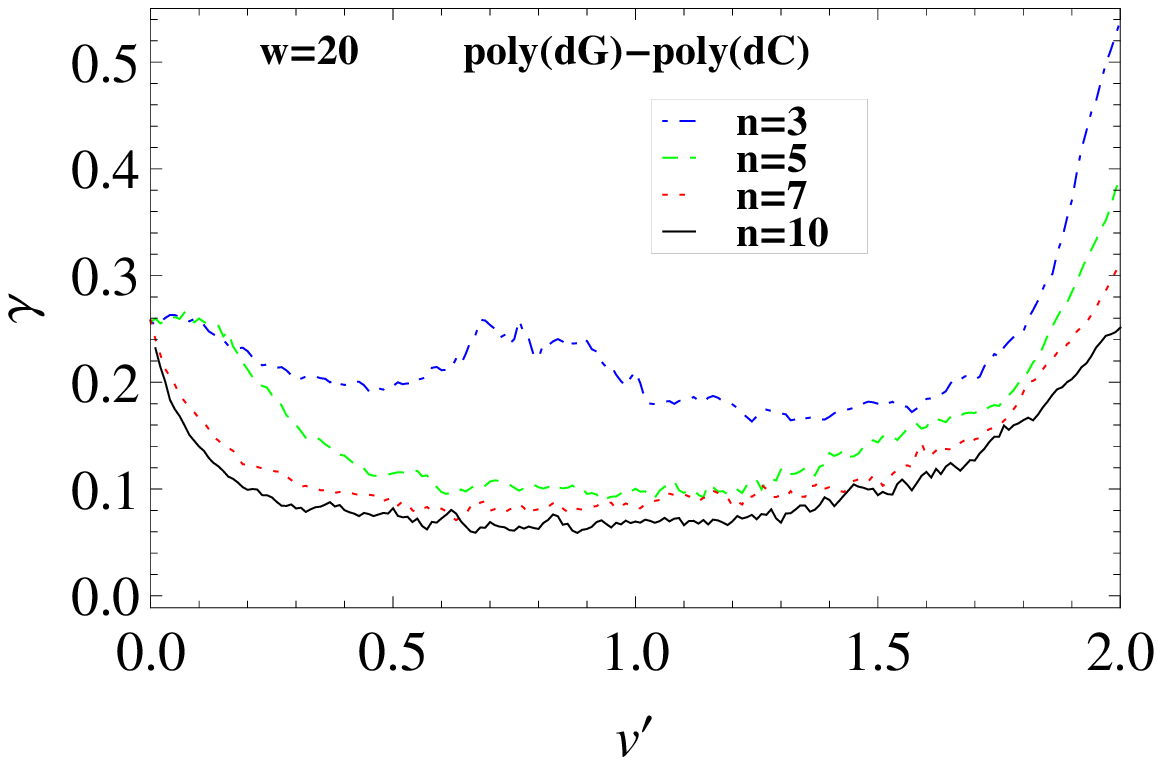}&
         \includegraphics[width=48mm, height=35mm]{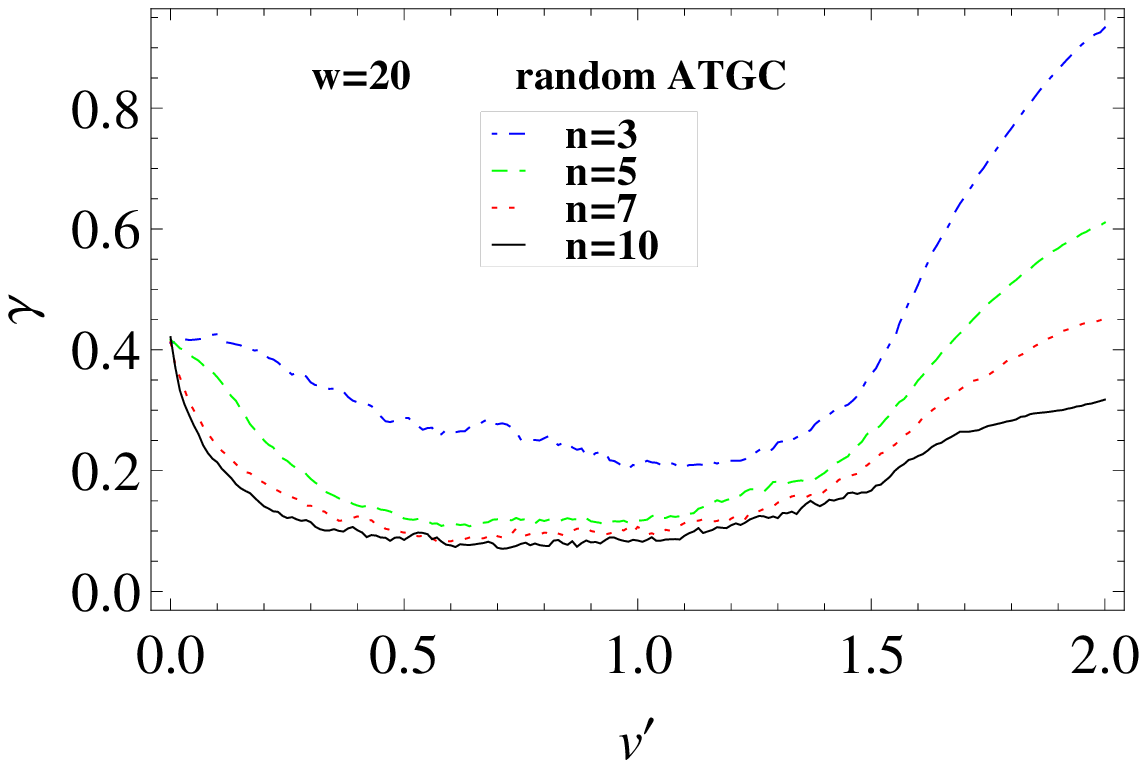}\\
              
  \end{tabular}

\caption{(Color online). Lyapunov exponent ($\gamma$) vs $v'$ for 
three DNA sequences at several disorder strengths (w), for 
four different values of n = 3, 5, 7, 10. $\gamma$ decreses with 
increasing values of n for all the sequences irrespective of 
disorder strength, though the features of localization curves 
are claerly distinguishable for different sequences. There is 
no distinct changes for the critical values of $v'$ (say, $v'_c$) 
which corresponds to the minima of $\gamma$ with n.}

\label{fig1}
\end{figure*}

    To obtain transmission probability $T(E)$ of electrons~\cite{datta1,datta2}
through DNA double-helix for this two-probe set up, we use the Green's function 
formalism. The single particle retarded Green's function operator representing 
the complete system {\it i.e.}, ds-DNA and two semi-infinite electrodes, at an 
energy $E$ can be written as $G^r=(E-H+i\eta)^{-1}$, where $\eta\rightarrow0^+$ and 
H is the Hamiltonian of the entire system. Using Fisher-Lee~\cite{datta1,datta2,fisher} 
relation the two terminal transmission probability is defined 
as $T(E)={\mbox {\rm Tr}} [\Gamma_L G^r \Gamma_R G^a]$, where 
$E$ being the incident electron energy and the trace is over the 
reduced Hilbert space spanned by the DNA molecule. The effective 
Green's functions can be expressed in the reduced Hilbert space 
in terms of the self-energies of the source and drain electrodes 
$G^r=[G^a]^\dagger=[E- H_{DNA}-\Sigma^r_S-\Sigma^r_D+i\eta]^{-1}$, 
where $\Sigma^{r(a)}_{S(D)}=H^\dagger_{\mbox{tun}} G^{r(a)}_{S(D)} 
H_{\mbox{tun}}$ and $\Gamma_{S(D)}=i[\Sigma^r_{S(D)}-\Sigma^a_{S(D)}]$, 
$G^{r(a)}_{S(D)}$ being the retarded (advanced) Green's function for 
the source (drain) electrodes. Here $\Sigma^r_{S(D)}$ and $\Sigma^a_{S(D)}$ 
are the retarded and advanced self-energies of the source (drain) 
electrodes due to its coupling with the DNA molecule. It can easily 
be shown that the coupling matrices $\Gamma_{S(D)}$ corresponding to 
the couplings of the DNA chain to the source (drain) electrodes 
$\Gamma_{S(D)}=-2~{\mbox{\rm Im}} (\Sigma^r_{S(D)})$. Whereas the 
self-energies are the sum of $\Sigma^r_{S(D)}$=$\Delta_{S(D)}$+i$\Lambda_{S(D)}$, 
$\Delta_{S(D)}$ being the real part of $\Sigma^r_{S(D)}$ corresponds to the 
shift of energy levels of DNA, and the imaginary part $\Lambda_{S(D)}$ is 
liable for the broadening of these levels.

 Considering linear transport regime, at absolute zero temperature, 
the two terminal Landauer conductance is given by $g=\frac{2e^2}{h}T(E_F)$, 
and the current passing through the DNA chain for an applied bias 
voltage V can be written as 
\begin{equation}
I(V)=\frac{2e}{h} \int^{E_F+eV/2}_{E_F-eV/2} T(E)dE~, 
\end{equation}

where $E_F$ being the Fermi energy. Here we have assumed that 
entire voltage drop occurs only at the boundaries of the conductor.

\section{Results and Discussions}

\begin{figure}[ht]
\centering

  \begin{tabular}{cc}
  
  \includegraphics[width=43mm, height=33mm]{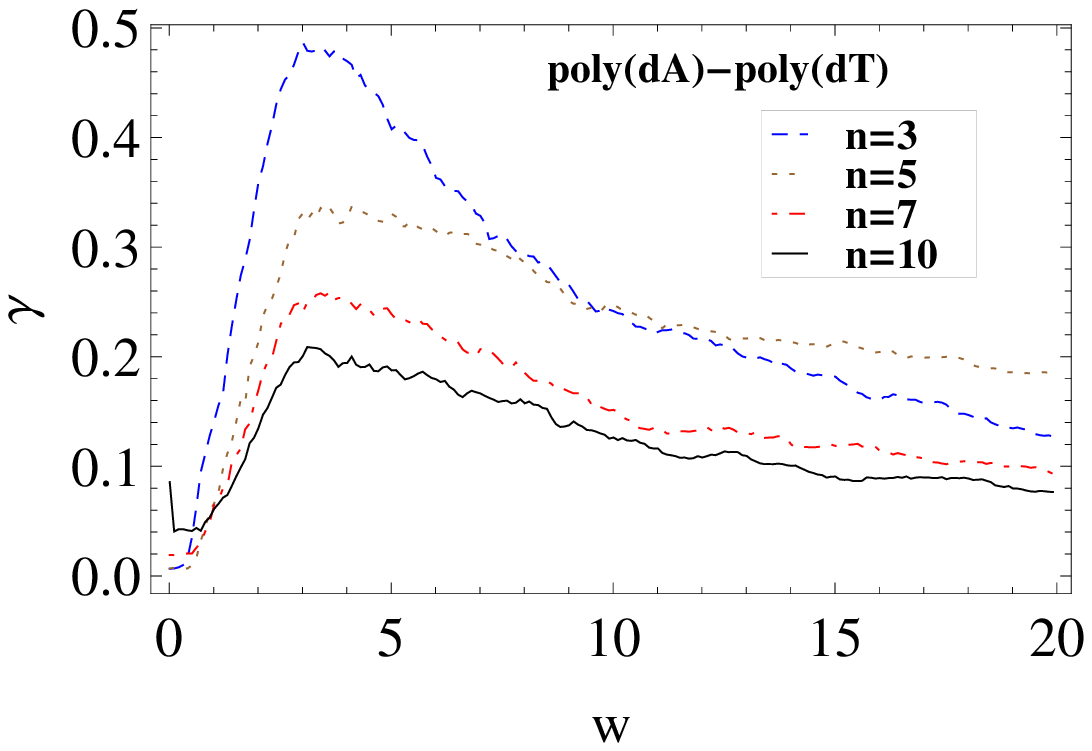}
   \includegraphics[width=41mm, height=32mm]{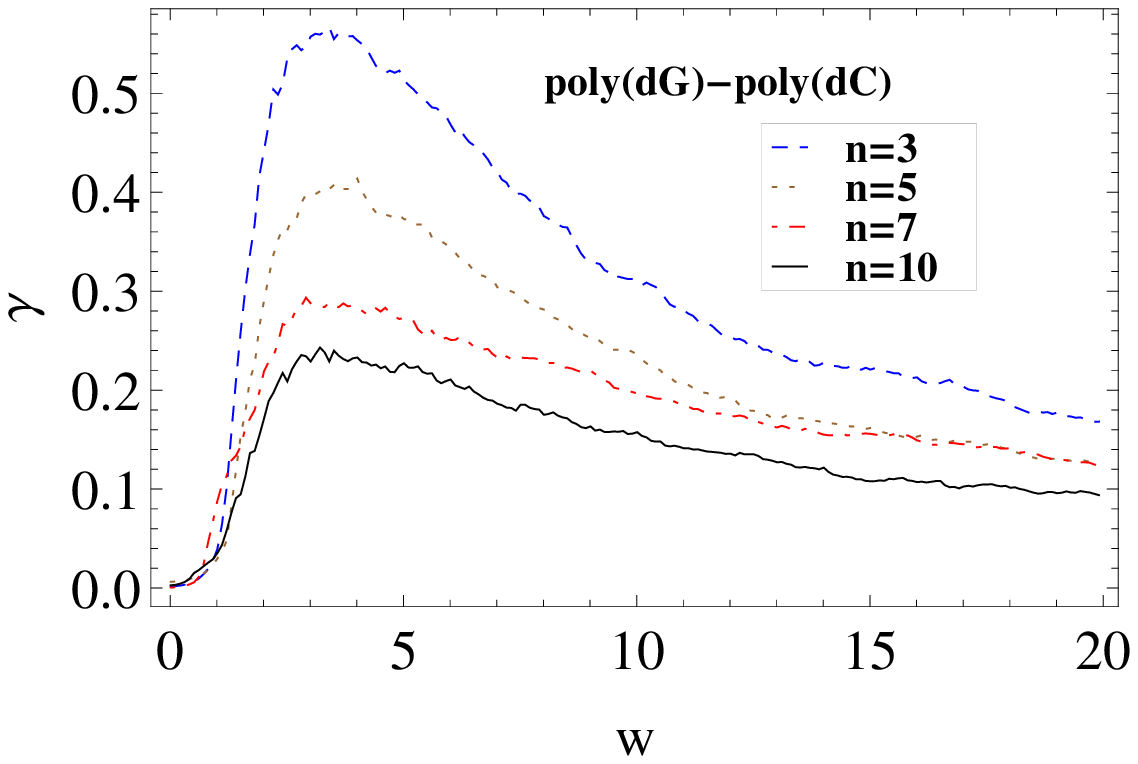}\\
   
    \includegraphics[width=44mm, height=34mm]{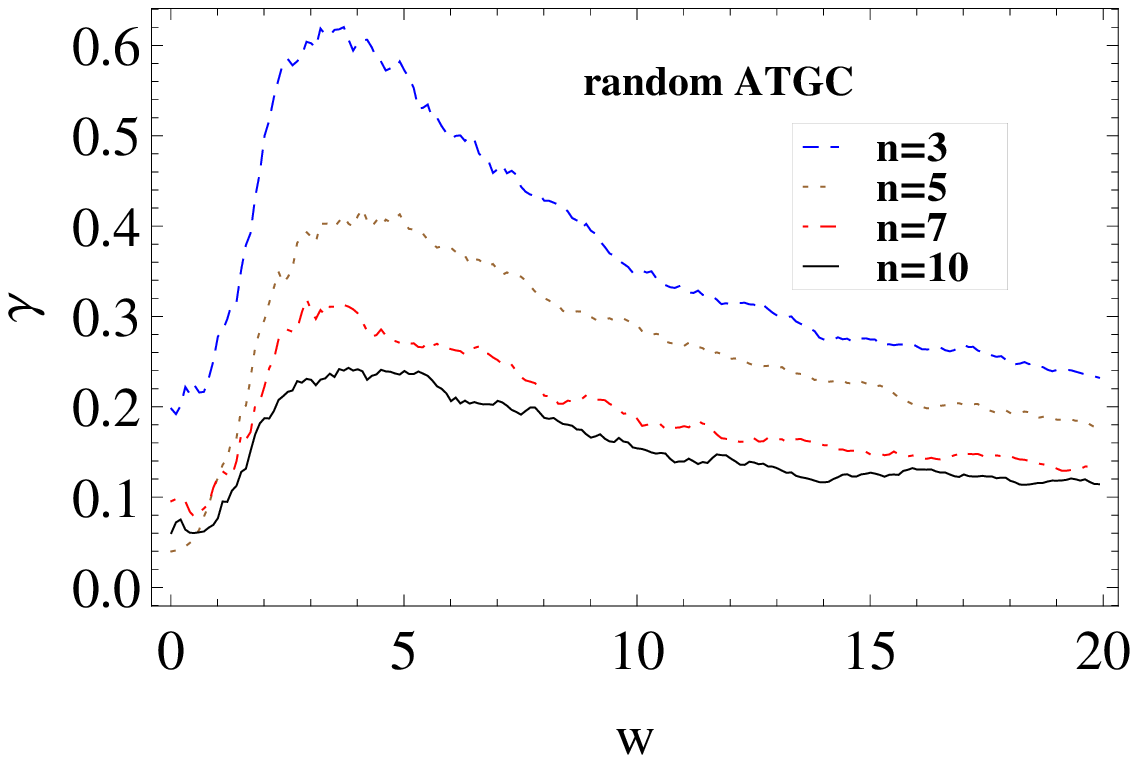}
    
  \end{tabular}

\caption{(Color online). Lyapunov exponent ($\gamma$) vs disorder (w) 
for three DNA sequences with $v'$=0.3 eV, for four different values 
of n = 3, 5, 7, 10. Unifrom behaviour of localization has been observed 
for all the sequences for whole range of disorder.}

\label{fig2}

\end{figure}

\begin{figure}[ht]

  \centering

  \begin{tabular}{cc}
  \includegraphics[width=42mm, height=32mm]{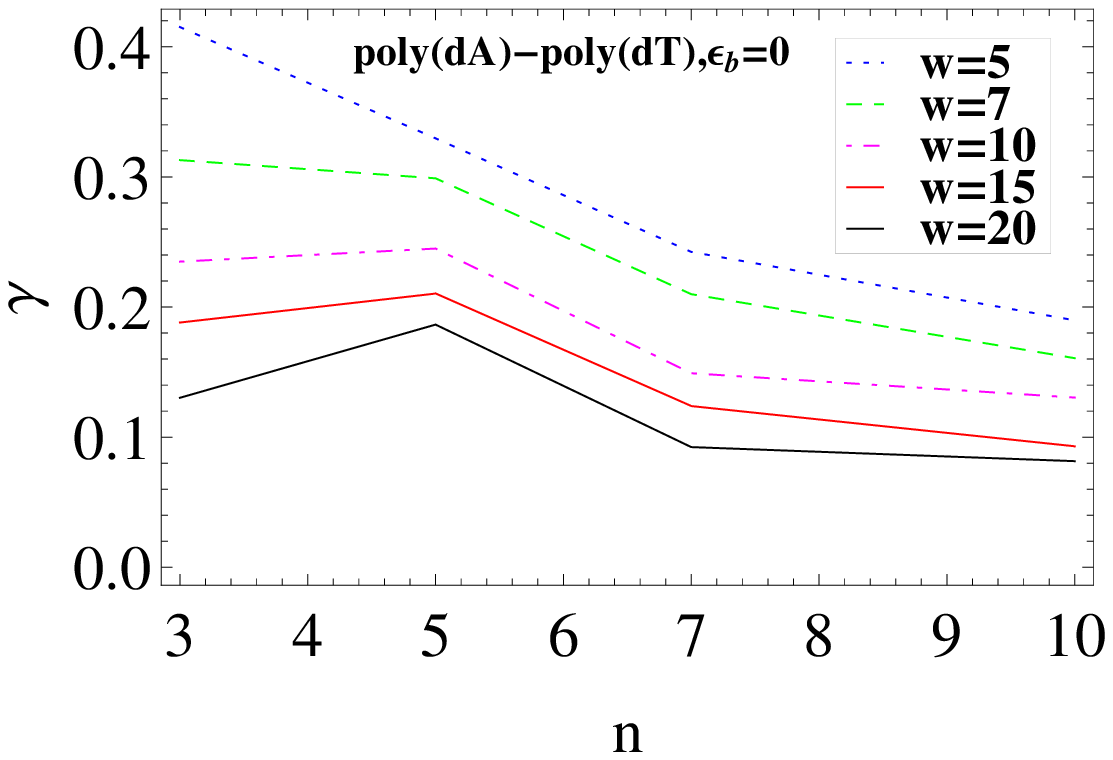}
   \includegraphics[width=42mm, height=32mm]{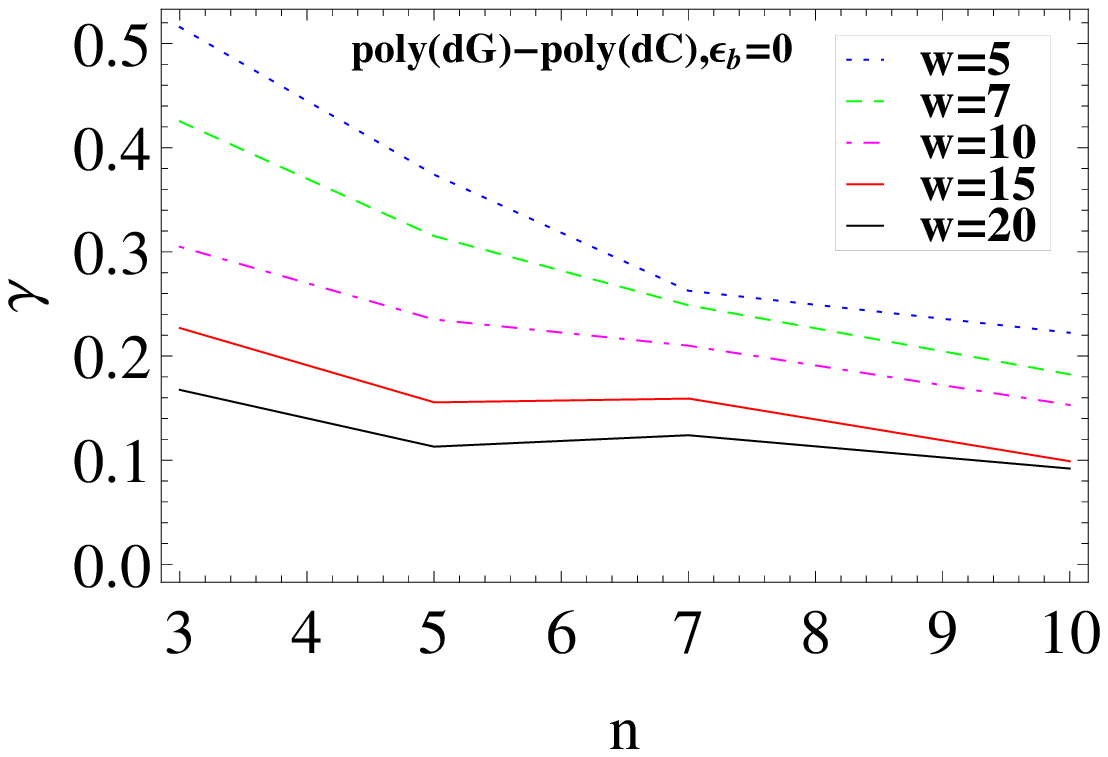}\\
   
    \includegraphics[width=44mm, height=34mm]{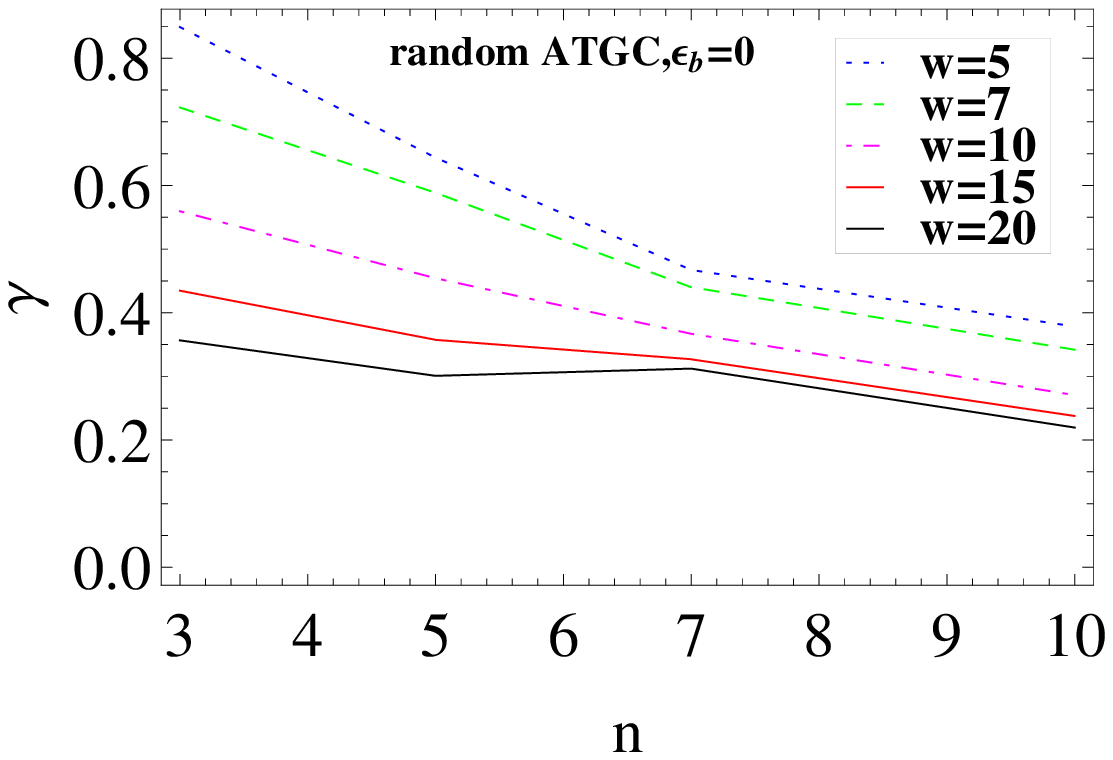}
  
  \end{tabular}

\caption{(Color online). Lyapunov exponent ($\gamma$) vs number of nucleotides 
within a pitch (n) for three DNA sequences with $v'$=0.3 eV, at different 
disorder strengths (w). Variation is quite uniform except for poly(dA)-poly(dT) 
sequence, a sharp peak is present there around n = 5 for heigher values of disorder, 
showing this may be the most localized configuration for that sequence.}

\label{fig3}

\end{figure}
  We first study the localization properties of the system by altering 
the number of bases in a given pitch of the helical structure. In order 
to do that we define localization length ($l$) from Lyapunov exponent 
($\gamma$)~\cite{ventra}, 
\begin{equation}
\gamma = 1/l = -\lim\limits_{L\to\infty}\frac{1}{L}<\ln(T(E))>~,
\end{equation}
where $L$ = length of the entire DNA chain in terms of basepairs,  
and $ <> $ denotes average over different disorder configurations. 
Though other distribution functions e.g., Gaussian and binary have 
been used to simulate experimental effects in previous studies~\cite{klotsa}, 
but we think it is appropriate to employ the most disordered case 
to simulate the actual experimental complications where the on-site 
energies of backbones $\epsilon_b$ to be randomly distributed within 
the range [$\bar\epsilon_b$-w/2, $\bar\epsilon_b$+w/2], where 
$\bar\epsilon_b$ is the average backbone site energy and w represents 
the backbone disorder strength. For the purpose of numerical 
investigation the on-site energies of the nucleotides are chosen as the  
ionization potentials of the respective bases, i.e., $\epsilon_G=-0.56 eV$, 
$\epsilon_A=-0.07 eV$, $\epsilon_C= 0.56 eV$, $\epsilon_T= 0.83 eV$. 
The intrastrand hopping integrals between identical nucleotides are taken as 
$t=0.35eV$ while those between different nucleotides are taken as $t=0.17 eV$.
We take interstrand hopping parameter to be $v=0.3 eV$. We emphasize 
that in case of the extended ladder model~\cite{paez, wells}, diagonal 
hopping between different nucleotides are also taken into account. But 
as in our case no diagonal hopping being considered, we compensate this 
by taking a quite larger value of interstrand hopping parameter $v$. Now 
as all the nucleotides are connected with sugar-phosphate backbones by 
identical C-N bonds, we take the hopping parameter between a base and 
corresponding backbone site same for all $t_b=0.7 eV$~\cite{cuni}. 
The parameters used here are the same as those used in~\cite{guo} which 
are consistent with {\it ab initio} calculations~\cite{voit,yan,senth}. 
For interstrand hopping $v'$ between nucleotides of adjacent pitches we 
follow Ref~\cite{sourav}. Nevertheless, we want to mention that  
choice of the tight-binding parameters is not unique and several 
parameter sets have been proposed in the existing literature~\cite{roche}.

\begin{figure*}[ht]

  \centering

  \begin{tabular}{ccc}

  \includegraphics[width=48mm, height=35mm]{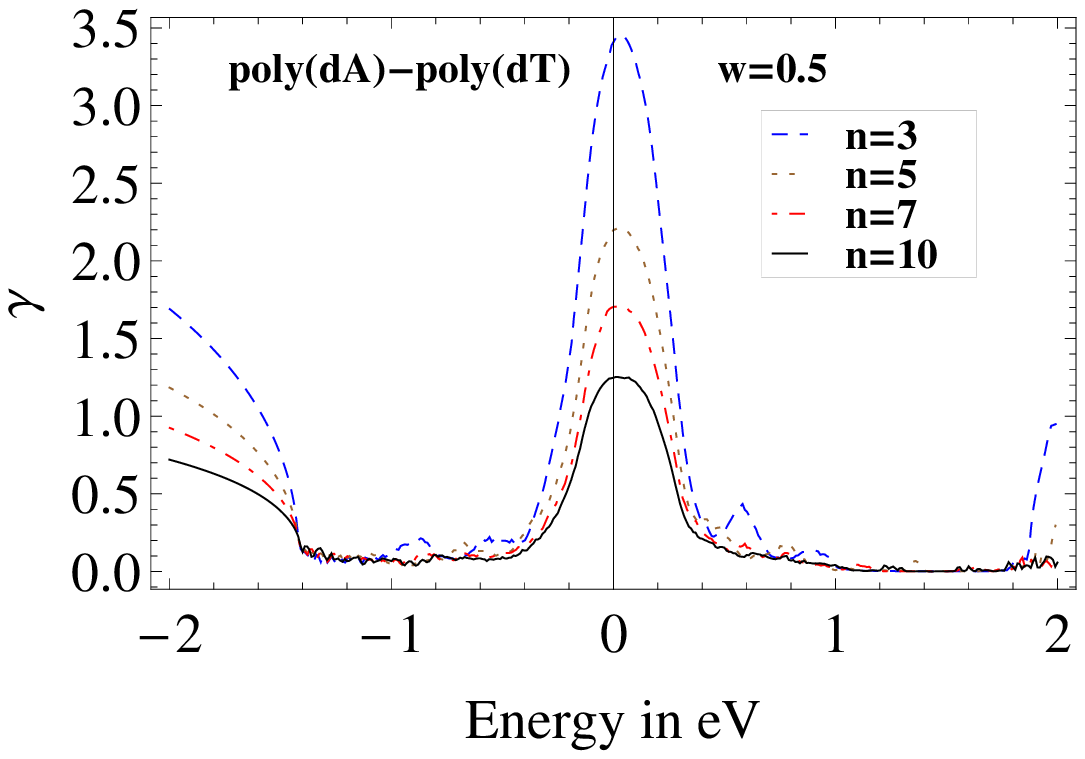}&
   \includegraphics[width=48mm, height=34mm]{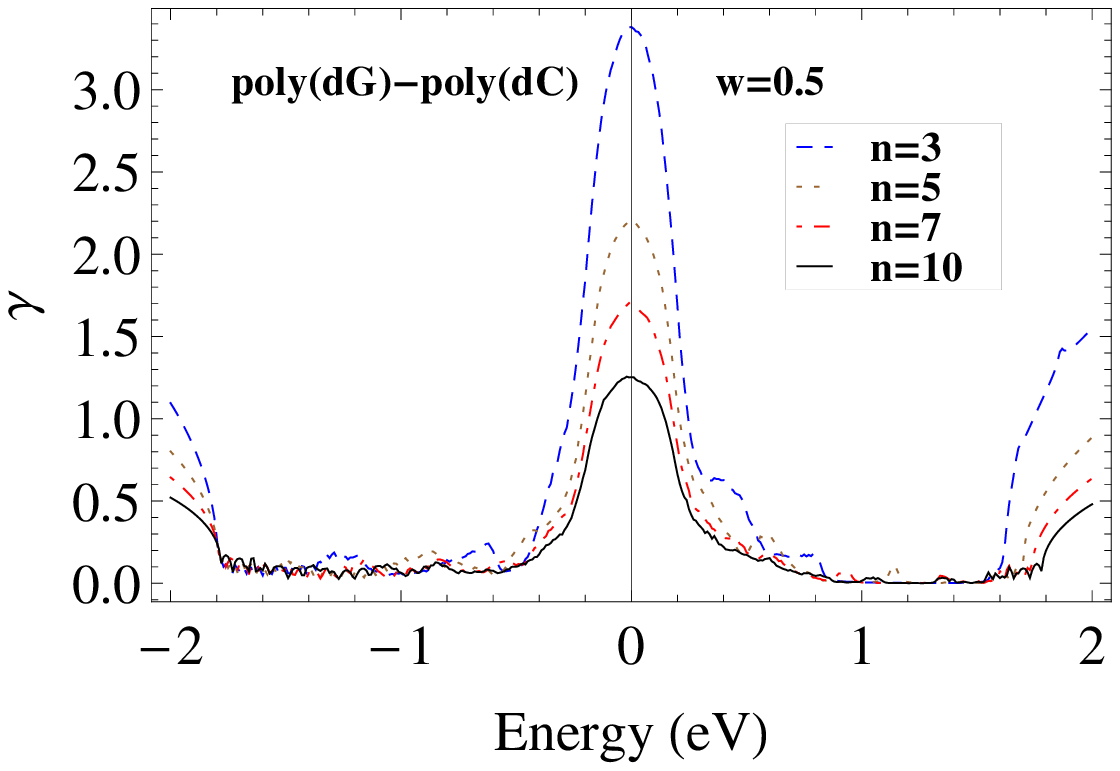}&
    \includegraphics[width=48mm, height=35mm]{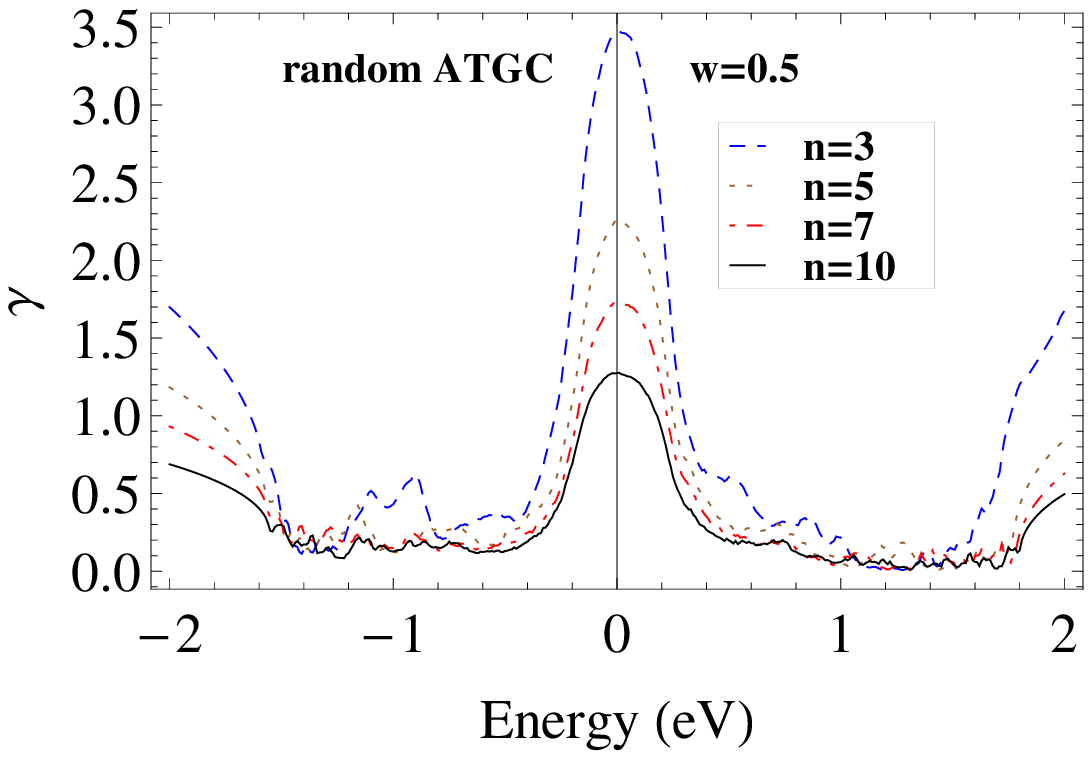}\\
    
   \includegraphics[width=48mm, height=35mm]{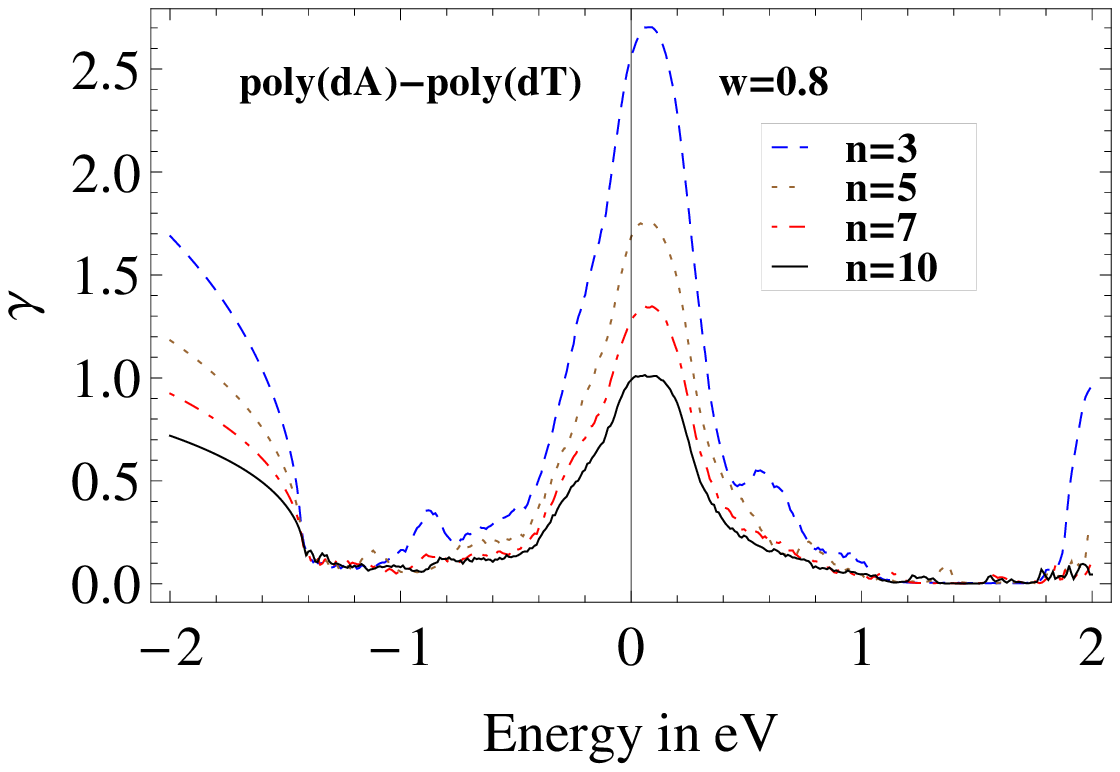}&
    \includegraphics[width=48mm, height=35mm]{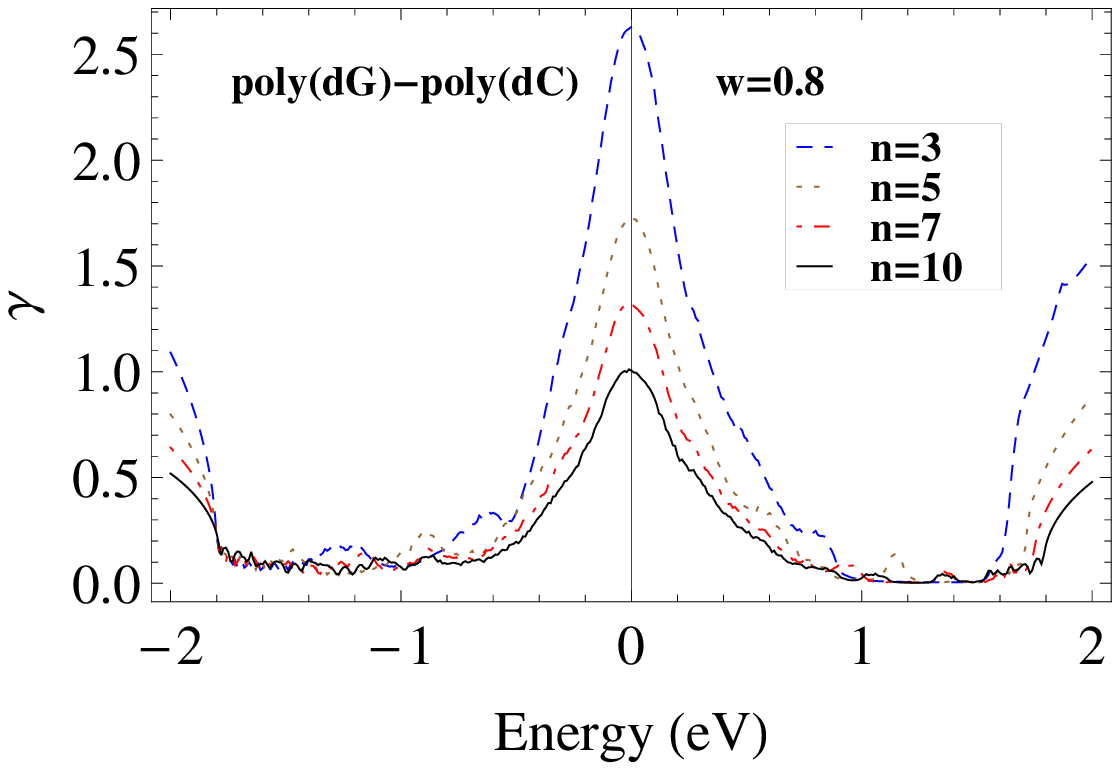}&
     \includegraphics[width=48mm, height=35mm]{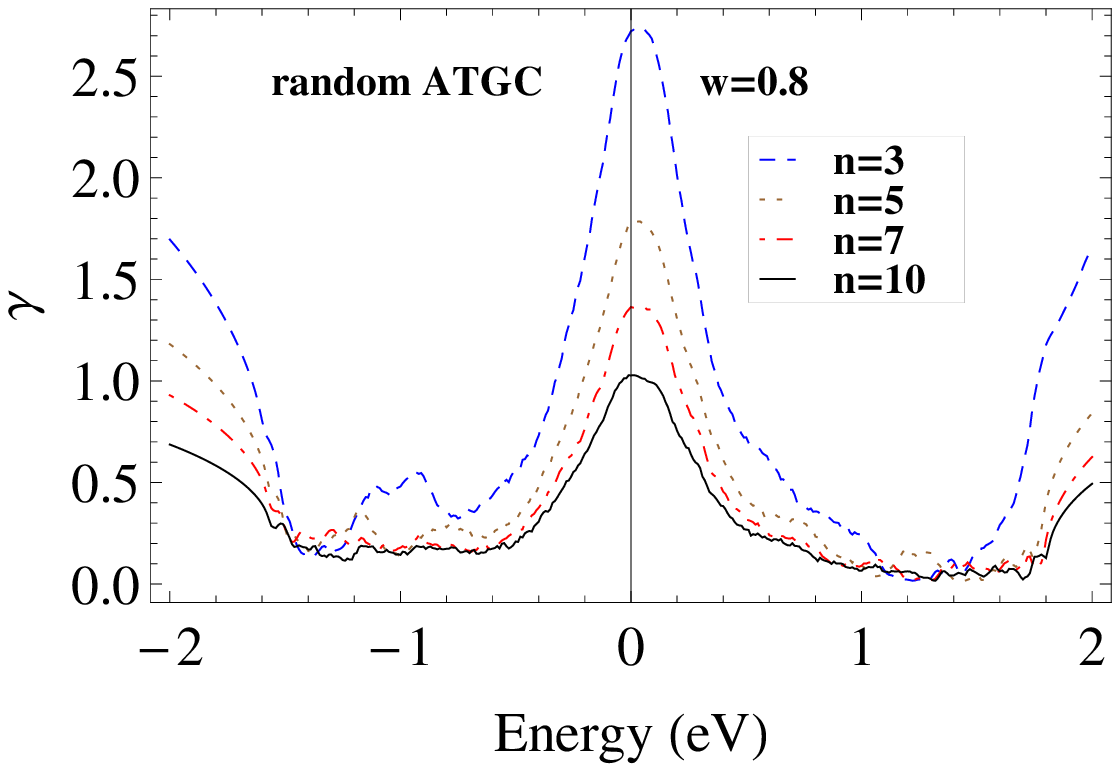}\\

    \includegraphics[width=48mm, height=35mm]{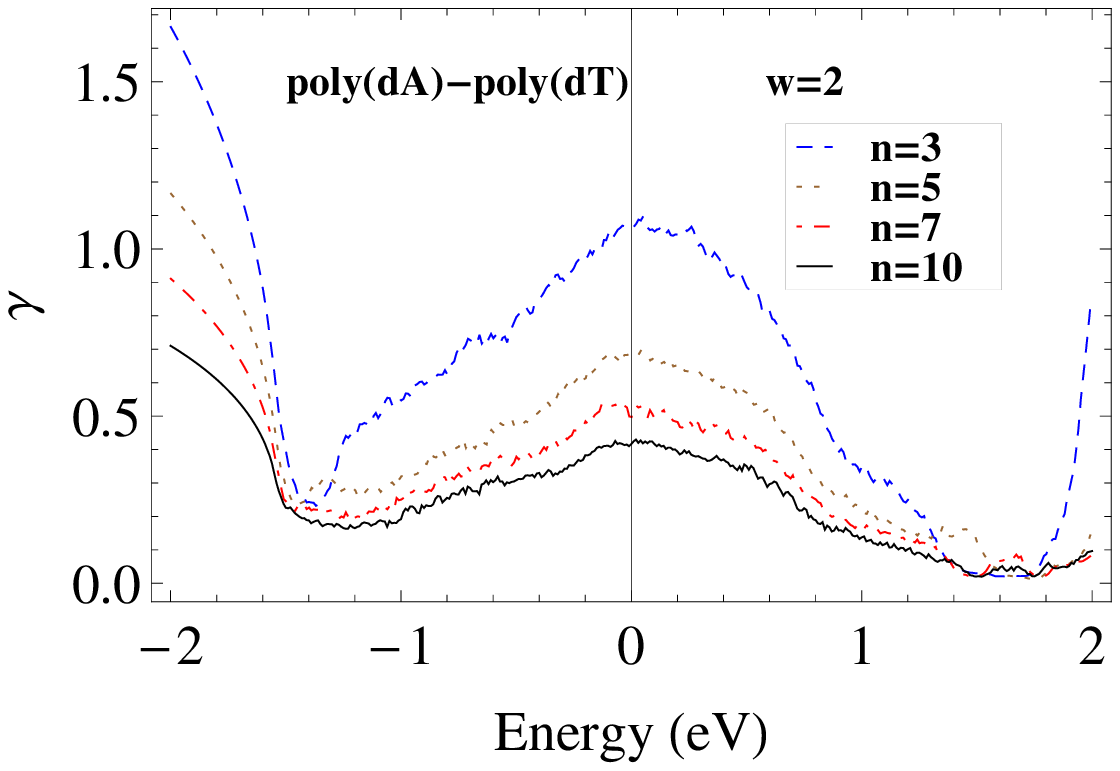}&
     \includegraphics[width=48mm, height=35mm]{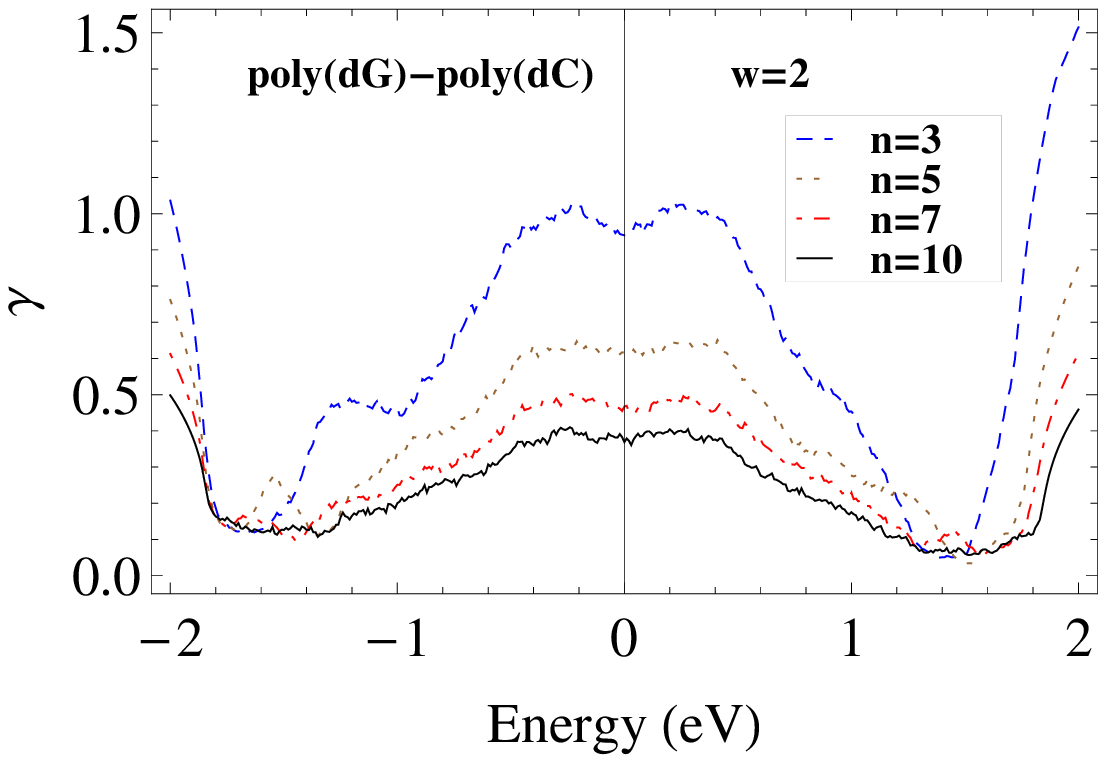}&
      \includegraphics[width=48mm, height=35mm]{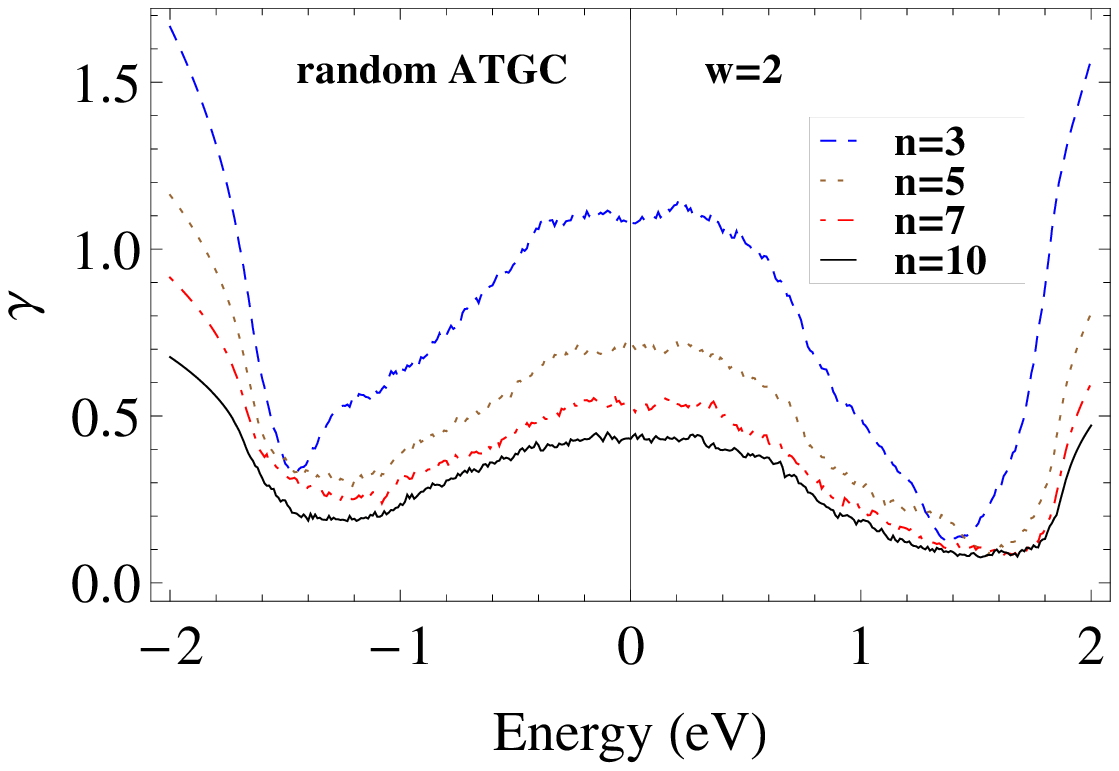}\\

     \includegraphics[width=48mm, height=35mm]{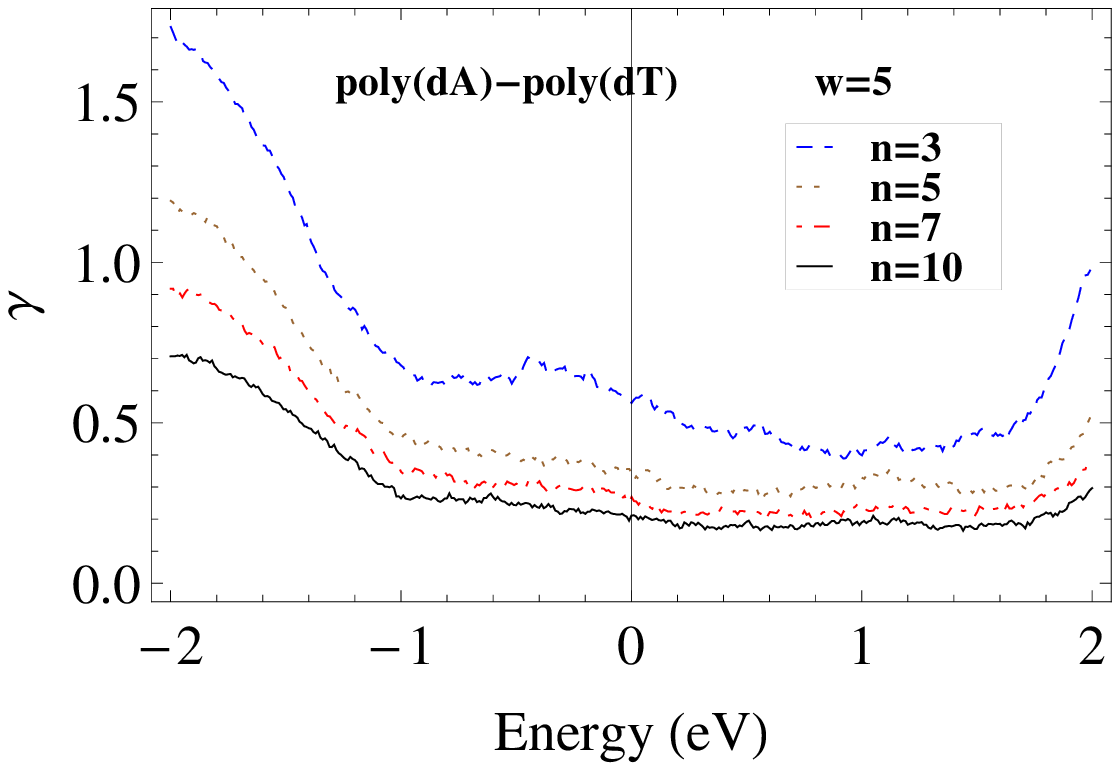}&
      \includegraphics[width=48mm, height=35mm]{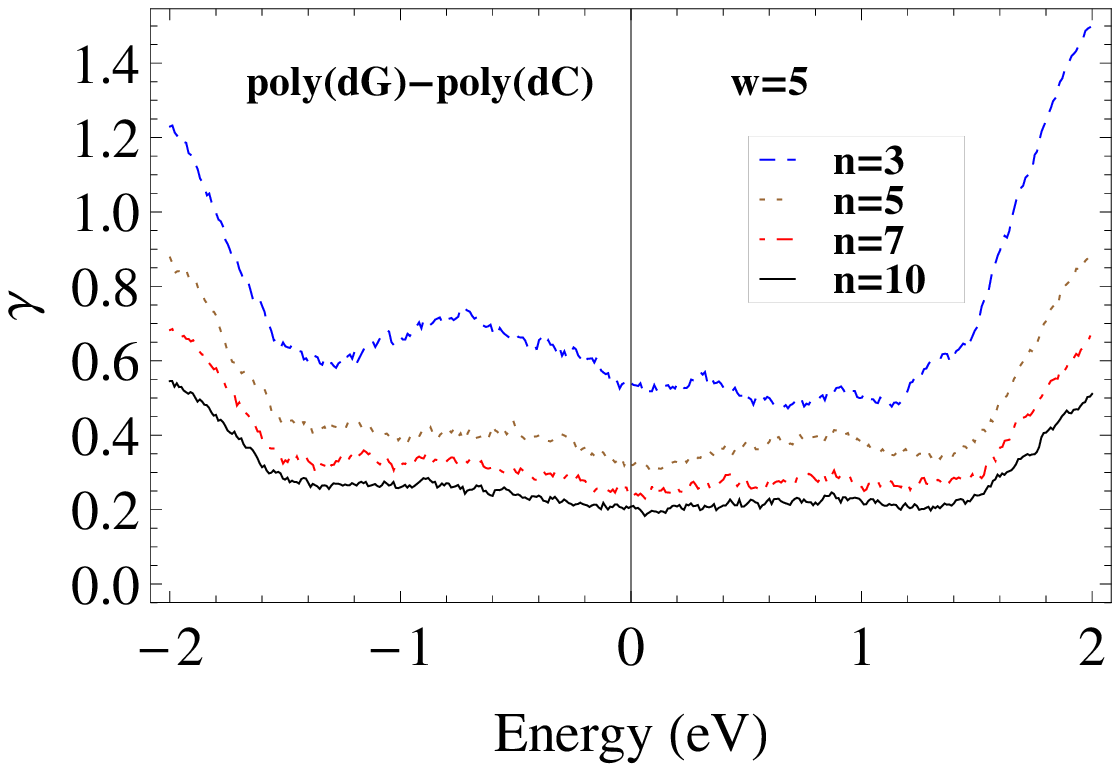}&
       \includegraphics[width=48mm, height=35mm]{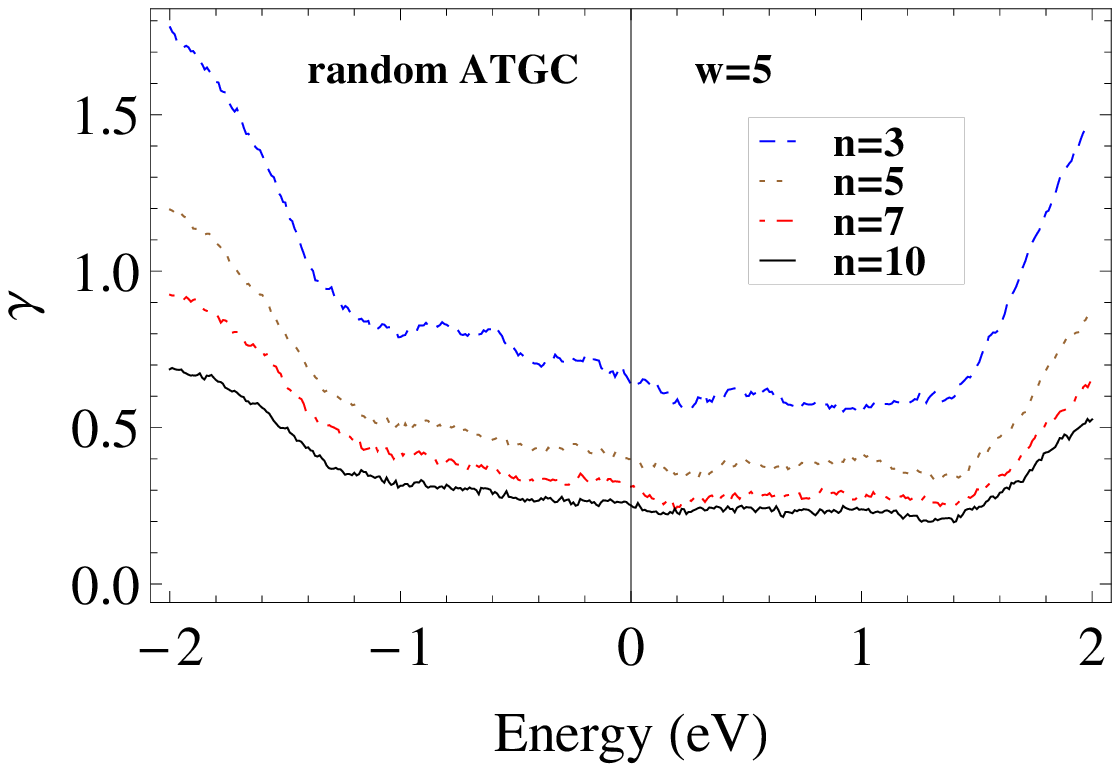}\\

      \includegraphics[width=48mm, height=35mm]{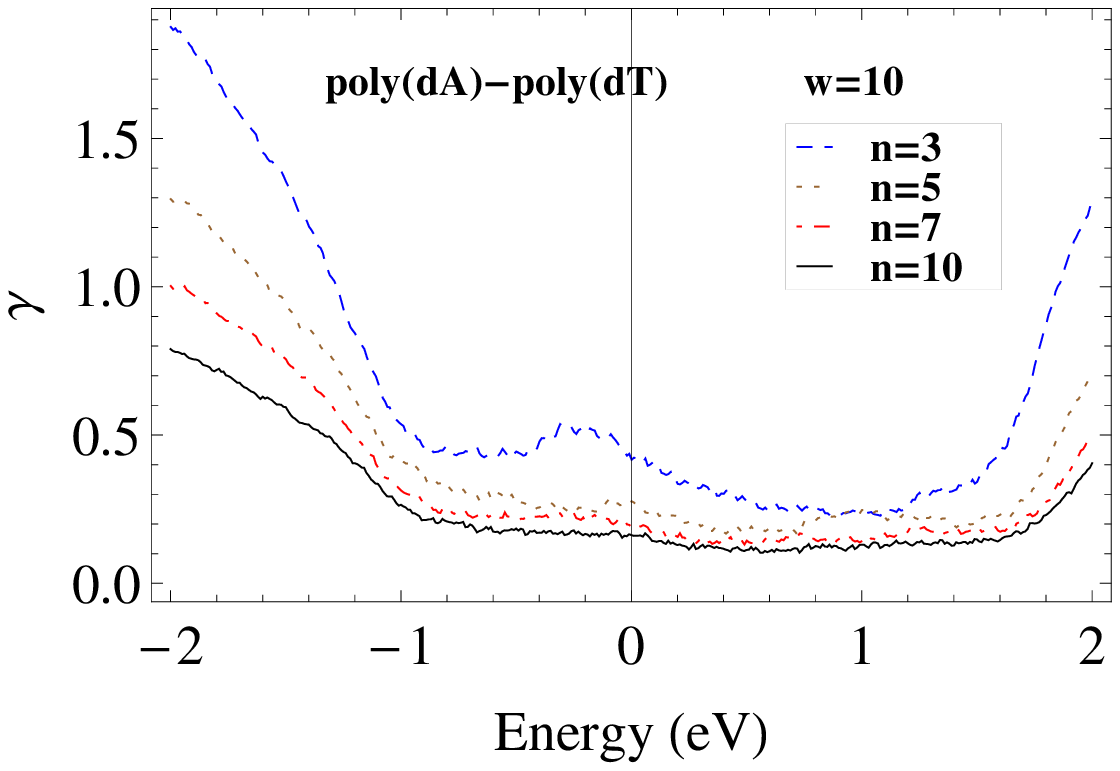}&
       \includegraphics[width=48mm, height=35mm]{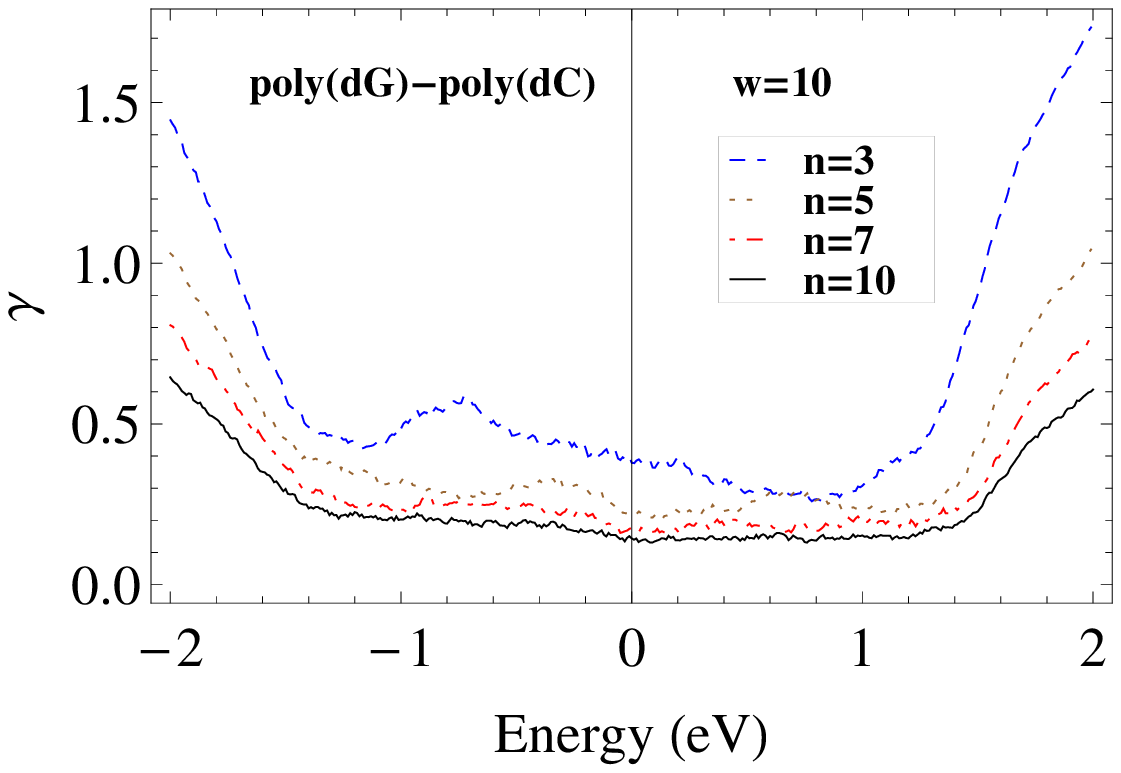}&
        \includegraphics[width=48mm, height=35mm]{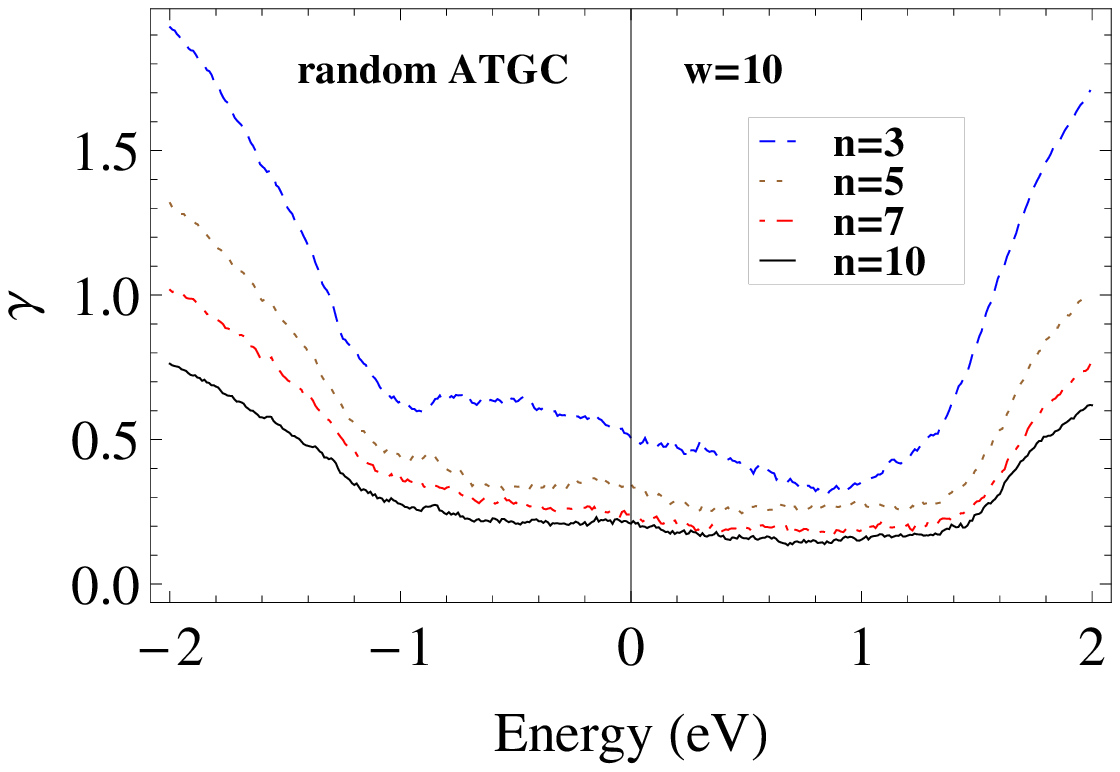}\\

  \end{tabular}
\caption{(Color online). Variation of Lyapunov exponent $\gamma$ with 
energy (E) for three DNA sequences with $v'$=0.3 eV, for different 
values of n. Effect of conformation (n) is stronger at the centre of band 
for low disorder (w) values, then it shifts towards the band edges 
for strong disorder.}

\label{fig4}

\end{figure*}
%It was previously shown that dispersion relation can be obtained 
%analytically for this model and the highest occupied molecular orbital 
%(HOMO) ($E_{-}$) and the lowest unoccupied molecular orbital (LUMO) 
%($E_{+}$) can be expressed as $E_{\pm}=\epsilon+2t\cos (k)\pm\sqrt
%{v^2+v'^2+2vv'\cos (nk)}$, which are separated by an energy gap 
%$E_g=2 \sqrt{v^2+v'^2+2vv'\cos (nk)}$~[Me]. For a fixed $v$ and 
%$v'$, energy gap ($E_g$) explicitly depends on the number of bases 
%($n$) in a given pitch of the helix .

 In Fig.~\ref{fig1} we have plotted the variation of inverse 
localization length ($\gamma$) for three sequences with $v'$ 
(which accounts for the helicity of DNA) at different values 
of n {\it i.e.}, number of nitrogen bases within a pitch of 
the helix, for various values of backbone disorder degree (w).
It is clear that all the curves have the same general shape 
for the periodic as well as the random DNA sequences and the 
variation of $\gamma$ with $v'$ is not monotonic. There exists 
a flat minima in these curves which indicates that at this point 
system is maximally extended. Now as we vary n (whatever be the 
disorder strength w is), $\gamma$ decreases, which indicates that 
system is less localized and effects of environmental fluctuations 
also becoming weaker. This behaviour can be explained easily, as we 
increase n we are allowing more channels for conduction between two 
adjacent pitches. As n increases, an electron can eventually hops 
from one pitch to the next, galloping other nucleotides in that pitch. 
With increasing n, the length of this gallop also increases {\it i.e.}, 
an electron gets the path to bypass more number nucleotides as it 
move along the DNA chain. Because of this the effective length become 
shorter for an electron and it feels less disorder. Hence, first due 
to helical symmetry system become less localized and then due to 
conformation (n) it gets more and more extended. So, at this configuration 
system is hardly effected by external disturbances. This information can 
help to perform experiments on DNA in more easier way and reproducible 
results can be generated which is a challenging task for a long time. 

\begin{figure}[ht]

  \centering

 \begin{tabular}{cc}

   \includegraphics[width=40mm, height=30mm]{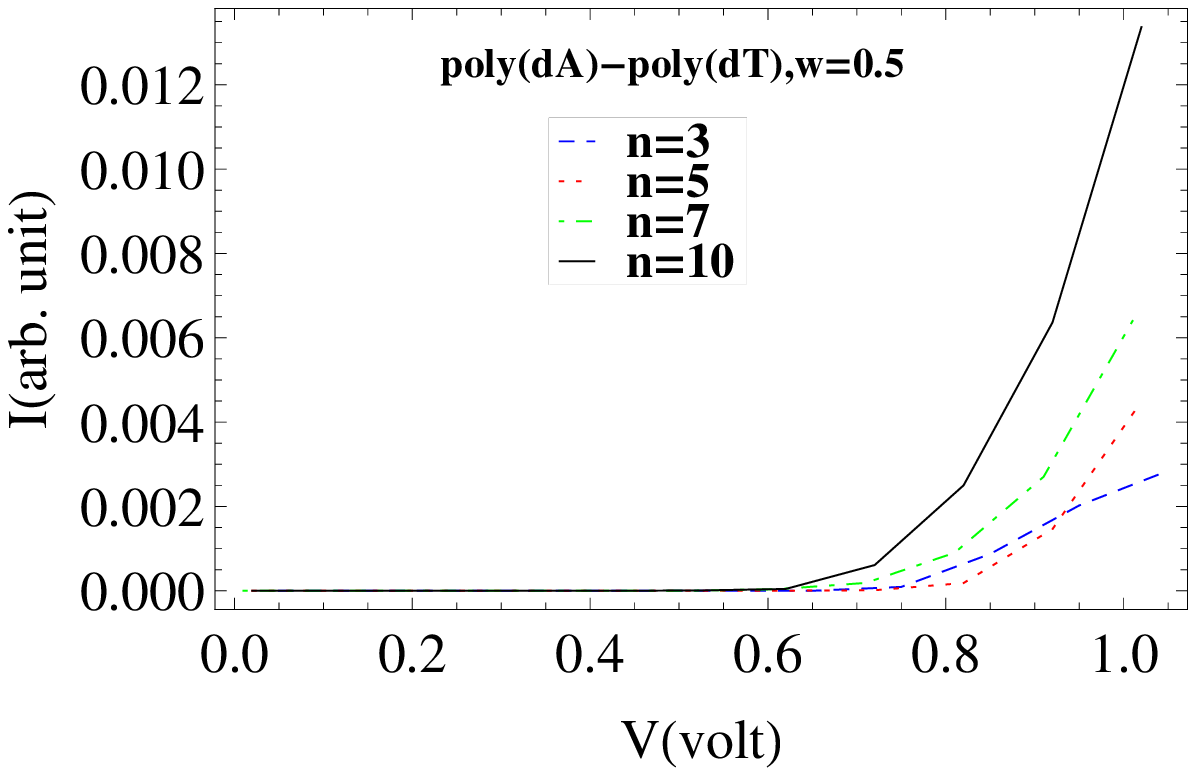}&
   \includegraphics[width=40mm, height=30mm]{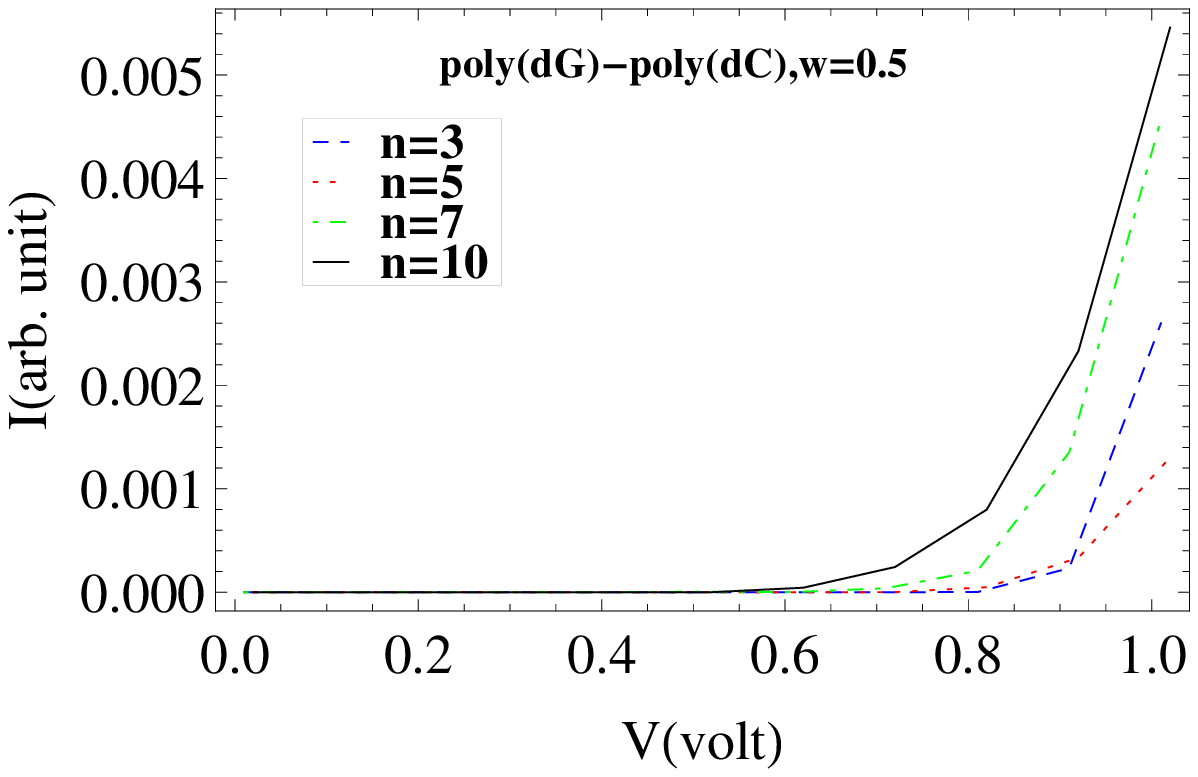}\\

   \includegraphics[width=40mm, height=30mm]{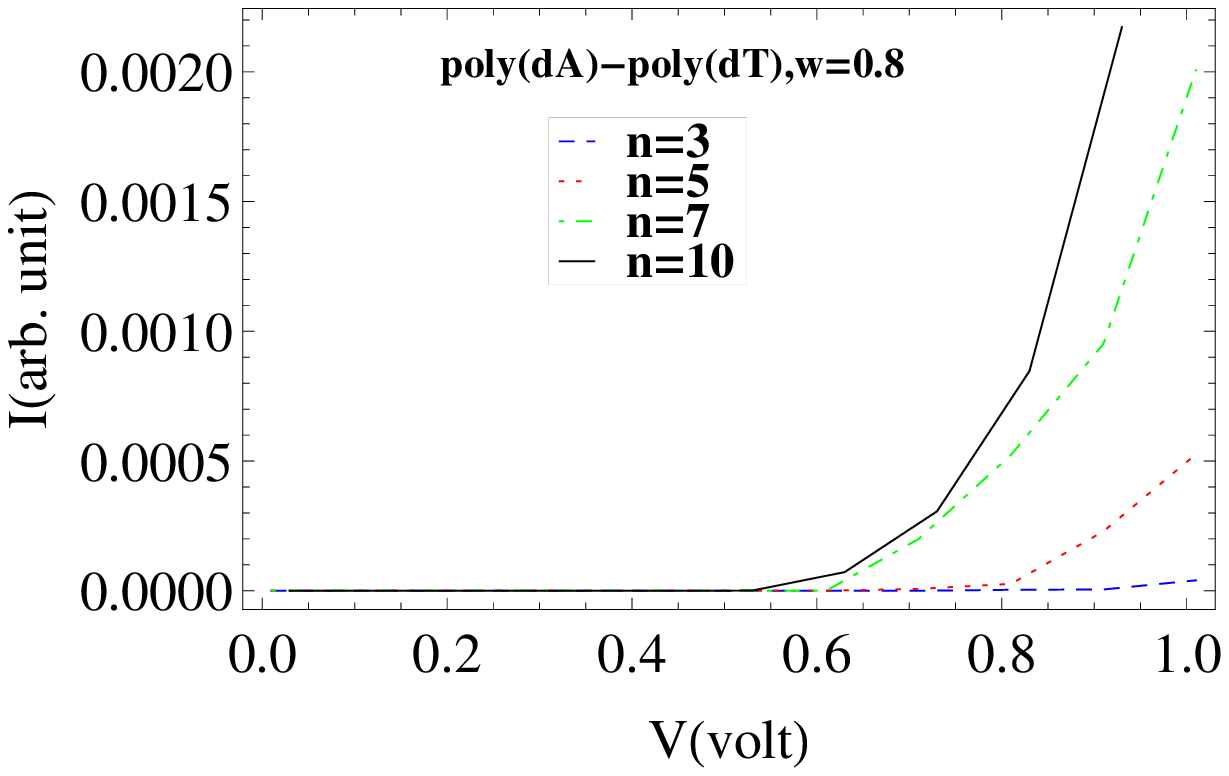}&
    \includegraphics[width=40mm, height=30mm]{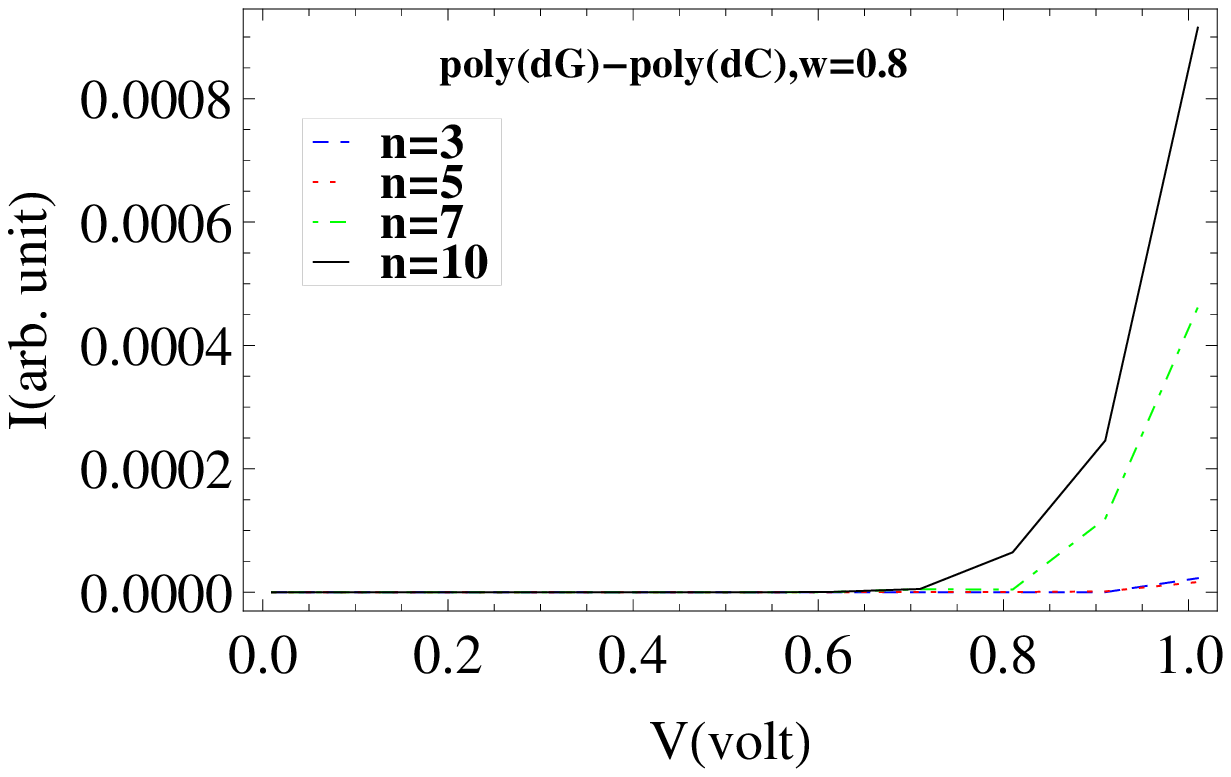}\\ 
   
    \includegraphics[width=40mm, height=30mm]{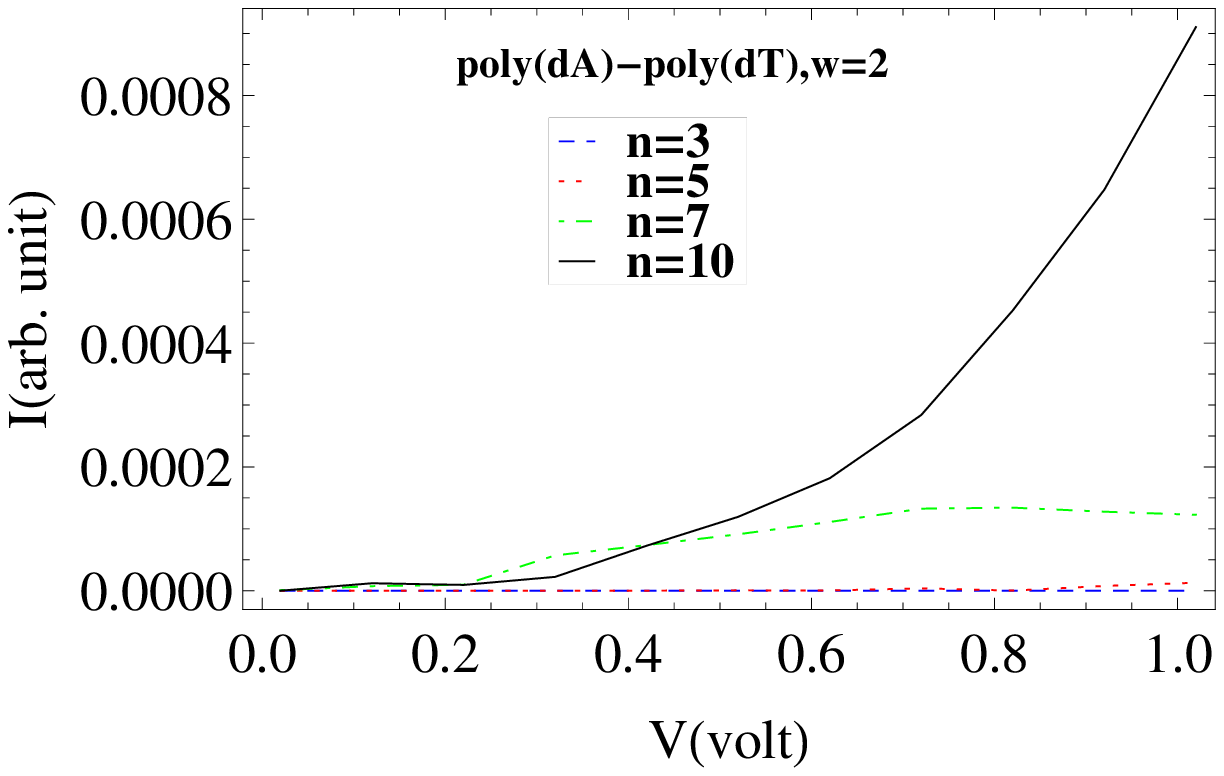}&
     \includegraphics[width=40mm, height=30mm]{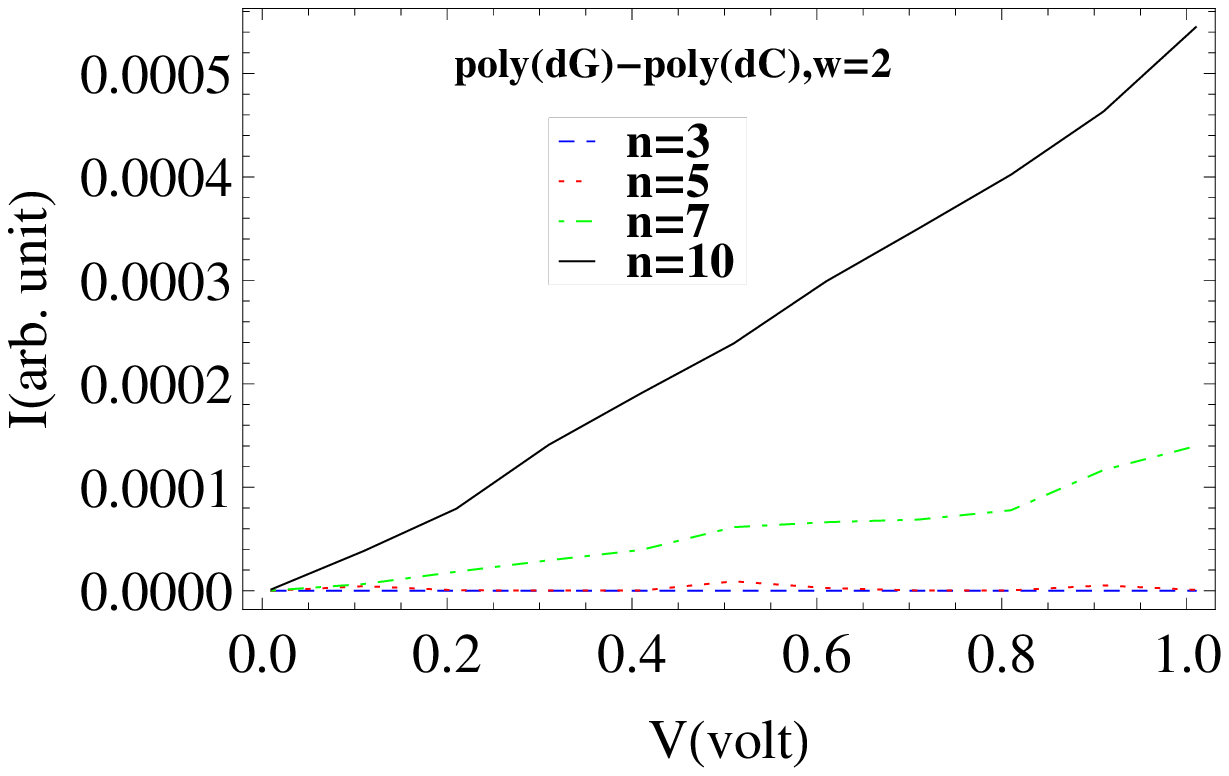}\\
     
     \includegraphics[width=40mm, height=30mm]{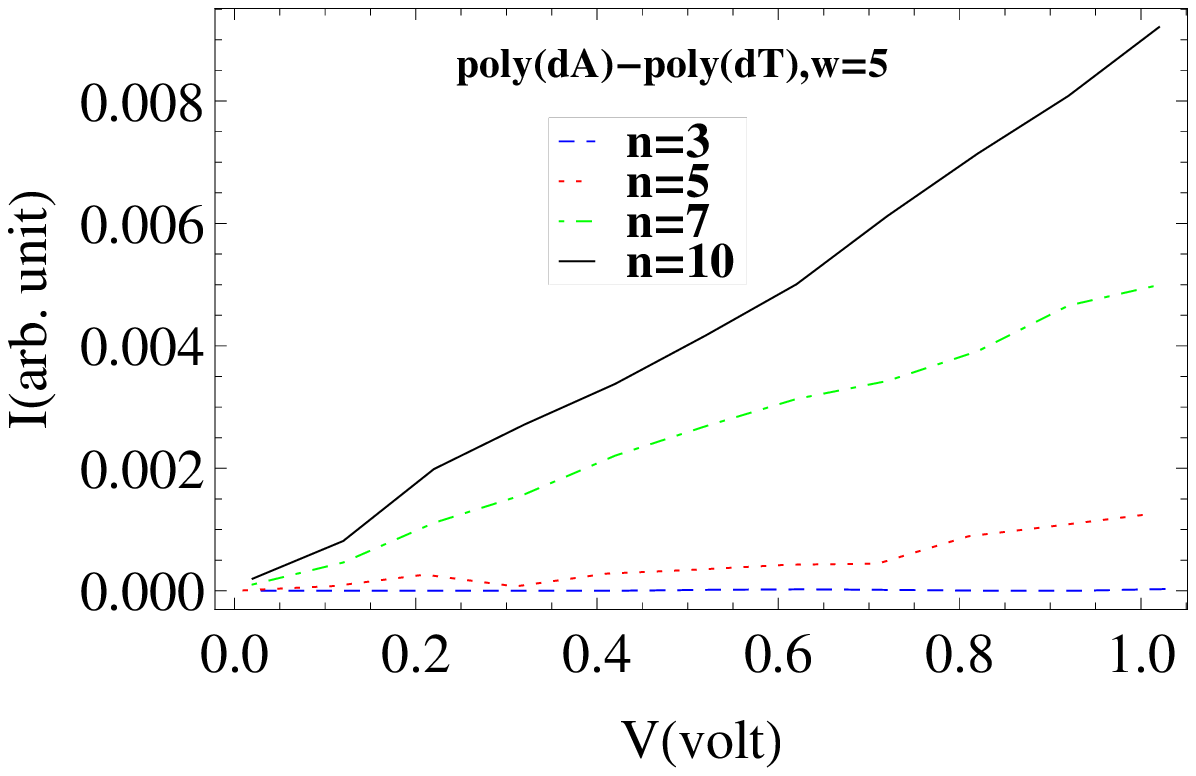}&
      \includegraphics[width=40mm, height=30mm]{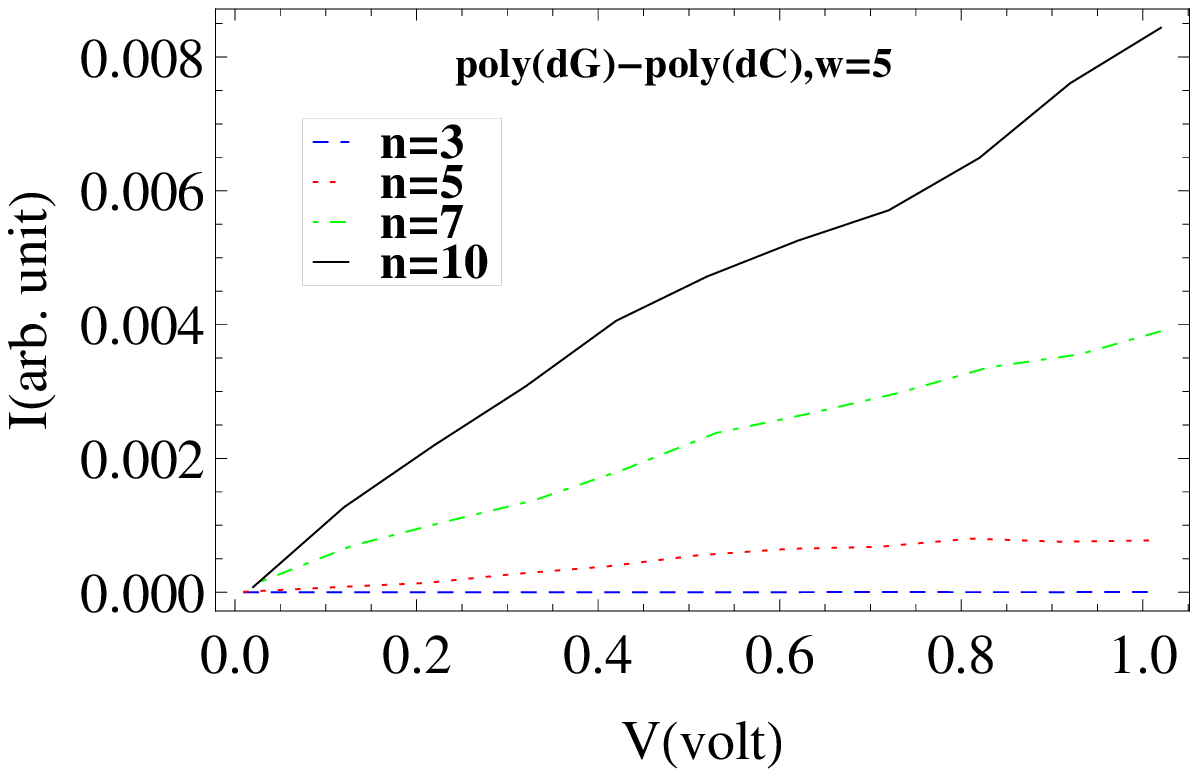}\\
    
      \includegraphics[width=40mm, height=30mm]{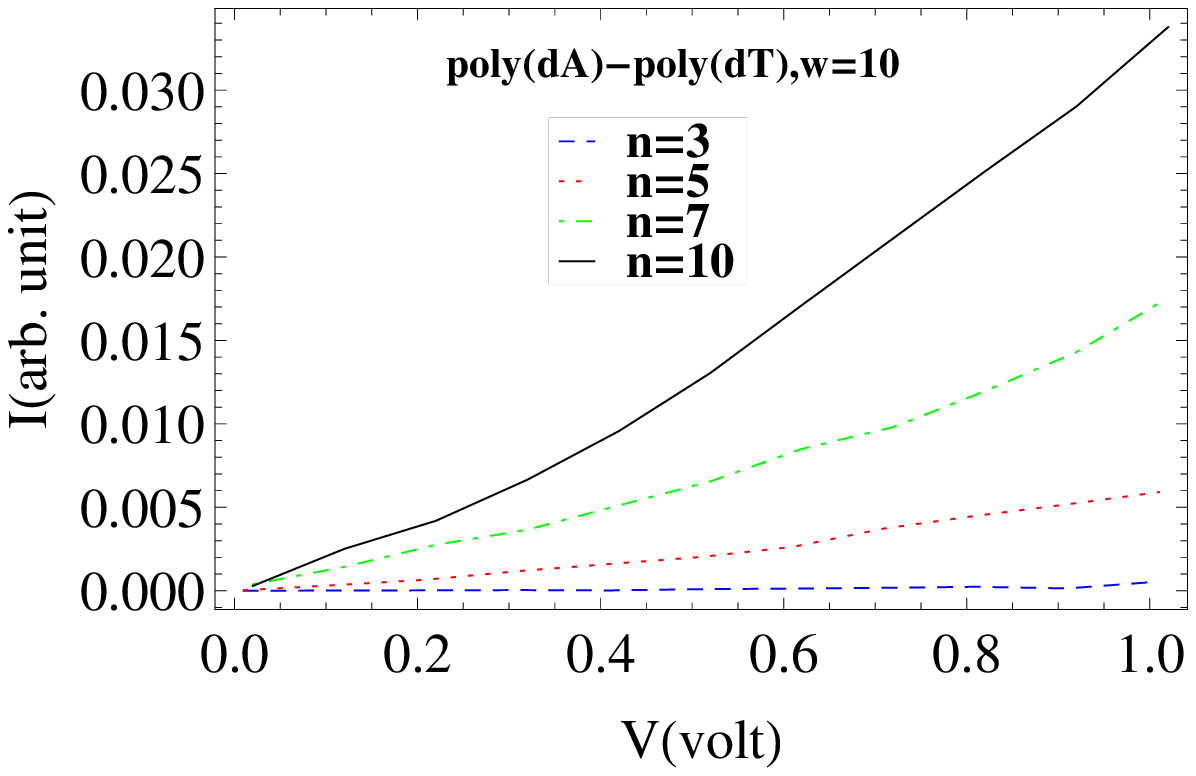}&
       \includegraphics[width=40mm, height=30mm]{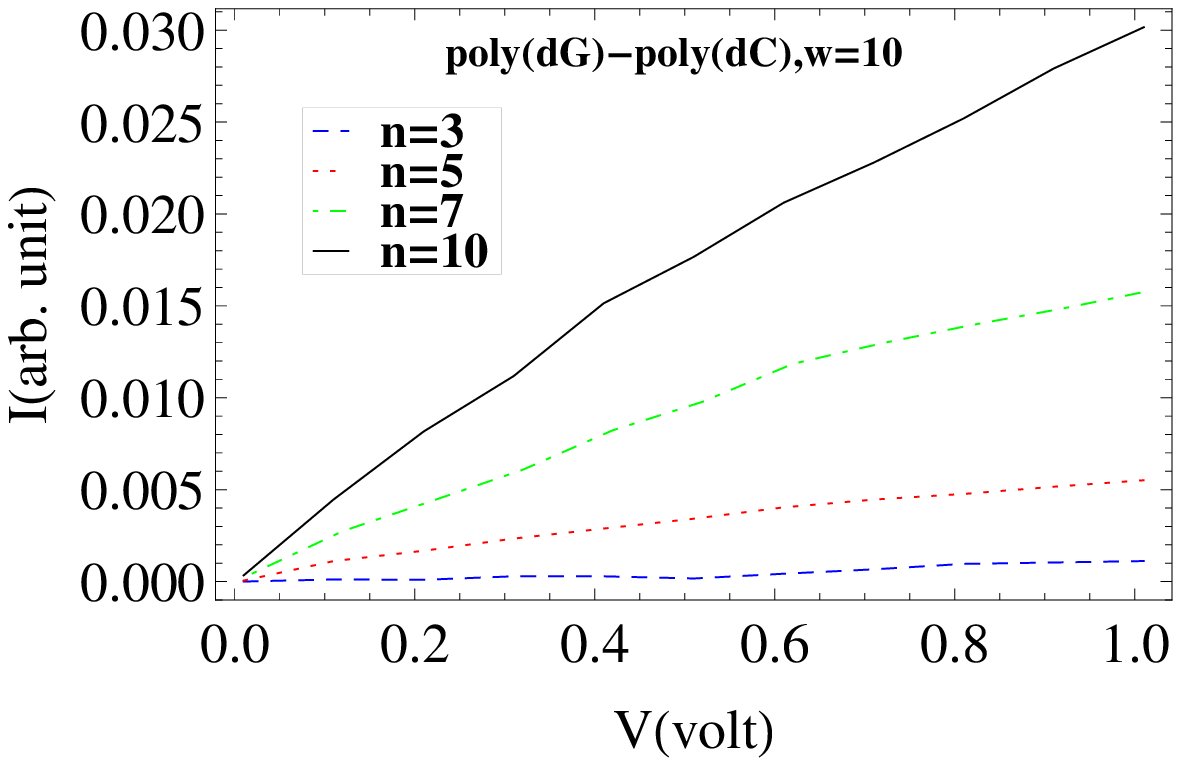}\\    
   
  \end{tabular}

\caption{(Color online). I-V response for two periodic sequences: poly(dA)-poly(dT) 
and poly(dG)-poly(dC) for five different disorder strength (w) at different 
values of n. For low disorder, cut off voltage reduces as we increase n, 
showing semiconducting behaviour. For strong disorder, current is considerably 
enhanced with increasing n giving a insulator to metallic transition.}

\label{fig5}

\end{figure}

 In Fig~\ref{fig2} we plot $\gamma$ vs. disorder strength (w), 
at a fixed value of $v'$=0.3 eV, for several values of n. Here 
also as we increase n, $\gamma$ decreases for all values of w. 
But the variation of $\gamma$ with w is also not unidirectional. 
$\gamma$ reaches a peak value for disorder strength within 3$<$w$<$4 
for all n values being considered. It signifies that at certain disorder 
level, localization length becomes minimum, which implies that at this 
point system is most effected by external disturbances. This typical 
behaviour of localization is due to backbone structure of DNA~\cite{guo}. 
The effect of variation of n is also less for low disorder compared 
to higher ones. The effect of conformation (n) is maximum when the 
system is at its most localized state (3$<$w$<$4).

In Fig~\ref{fig3} we show the variation of $\gamma$ with n. It also 
shows with increasing n, $\gamma$ decreases, except there is some 
different features around n=5 for one of the sequence (ploy(dA)-poly(dT)). 
Though the reason is not yet clearly understood but it seems that system 
may has a critical configuration, at which it does feel environmental effects 
most as we vary n. Because of that at n=5, $\gamma$ increases instead of 
decreasing, showing it is the most localized configuration under appreciable 
disorder. This behaviour is not present in the other sequences, which shows 
different localization behaviour depending on sequential variety. 

 We also investigate localization behaviour with energy. 
In Fig~\ref{fig4} we plot the variation of $\gamma$ with 
energy for different values of n. The same thing is also 
happened here, with increasing n, $\gamma$ decreases. Though 
the rate of decreasing is fast for small n (n=3, 5), then the 
variation of n become less effective for changes at higher 
values of n (n=7, 10). Effect is more prominent near centre of 
the bands for low disorder. As the disorder increases, effect 
of variation of n gradually delocalize towards the edges. At 
high disorder, variation is more sensitive around the edges of 
the band rather being at the centre.

In Fig.~\ref{fig5} we plot I-V characteristics for the two periodic 
sequences for several values of n. We set the temperature at 0 K. To 
minimize the contact effects we choose tunnelling parameter $\tau$ to 
be optimum $\it{i.e.}$, $\tau$=$\sqrt{t_{ij}\times t}$ between ds-DNA 
and the electrodes, where t is the hopping parameter for the electrodes
~\cite{macia}. %which we choose 4 eV and site energies of the semi-infinite 
%1D electrodes is set to 0 eV. 
It is clear that effect of n is less 
at low disorder which is obvious because at low disorder any path of 
charge conduction is equivalent as an electron feels almost no potential 
variation. As the disorder increases effect of n becomes more distinctive. 
For strong disorder there is substantial variation of potential at different 
sites and change in n gives an electron more number of shortcut pathways to 
move along the DNA chain. So, with increasing n, current is enhanced and 
the effect is sharp for high disorder values. For low disorder values 
cut-off voltage being reduced with increasing n, showing semiconductor-like 
transport. At high disorder for both the periodic sequences, current is 
considerably enhanced and almost linear response is observed at higher 
values of n, which indicates a transition from insulating to metallic phase. 
Our results are consistent with several experimental findings~\cite{fink, porath, storm}. 
  
\section{Concluding Remarks}
	
 Till now different models have been used to study transport properties 
of DNA but none of these has taken into account of helical symmetry 
which is a basic feature of DNA structure. Using twisted ladder model 
we first incorporate the helicity and then by varying the number of 
nucleotides within a pitch we try to model the conformational variation 
of DNA. Though some calculations are present in the literature~\cite{yega, 
song, maragakis} but investigation within tight-binding framework is lacking. 
We report that depending on helical symmetry and conformation, localization 
properties can change considerably. The effect of conformation is less when 
environmental complications are small and increases with it. We have two interesting 
results. First one is by incorporating helical symmetry and conformation we have 
been able to minimize the environmental effects to a great extent. It is clear 
from localization data that interplay of helical symmetry and conformation 
can provide some configurations where system is hardly disturbed by external 
agencies. If this information can be used correctly in experiments, 
we think the operation of such experiments would become less complicated. 
We investigated these properties in every aspect possible and it shows 
unambiguous variation with conformational changes. The second result is, 
in presence of helical symmetry, depending on the cooperative effect of 
backbone disorder and conformation system can undergo a transition from 
insulating to metallic phase as it is eminent from the I-V responses of 
periodic sequences for higher disorder values. Whereas for low disorder with 
increasing n, cut-off voltage being reduced for semiconducting response. In 
summary, we can say that conformal changes have prominent effects on charge 
transport properties of DNA as it shows that DNA can be found in three different 
phases e.g., insulating, semiconducting and metallic depending on the mutual variation 
of environmental fluctuations and conformation. We hope in near future our results 
will be tested experimentally to find exact effects of helical symmetry as well as 
conformation on transport properties of DNA.

\end{document}